\documentclass[aps,prl,superscriptaddress,notitlepage,nopacs,amsmath,amstex,amssymb,citeautoscript,longbibliography,floatfix,letter,twocolumn]{revtex4-2}
\usepackage[utf8]{inputenc,}
\usepackage[bookmarks=false,linkcolor=blue,urlcolor=blue,colorlinks,citecolor=blue]{hyperref}
\usepackage{amsmath}
\usepackage{amsfonts}
\usepackage{amssymb}
\usepackage{graphicx}
\usepackage[left=2cm,right=2cm,top=2cm,bottom=2cm]{geometry}
\usepackage[dvipsnames]{xcolor} 
\usepackage{fancyhdr} 
\usepackage{titlesec} 
\usepackage{amsmath,amsthm,amsfonts,amssymb}
\usepackage{graphicx, soul}
\usepackage{fancyhdr, color}
\usepackage{titlesec}
\usepackage{enumitem}
\usepackage{graphicx,amsmath,amssymb,color}
\usepackage{nicefrac}
\usepackage{multirow, makecell, ragged2e}
\usepackage[titletoc,title]{appendix}
\usepackage{bm}
\usepackage{float,lipsum}
\usepackage{algpseudocode}
\usepackage{algorithm,cancel,soul,ulem}
\usepackage{xspace}
\newcommand{\bra}[1]{\langle #1|}
\newcommand{\ket}[1]{|#1\rangle}
\newcommand{\be}{\begin{equation}}
\newcommand{\ee}{\end{equation}}
\newcommand{\beqr}{\begin{eqnarray}}
\newcommand{\eeqr}{\end{eqnarray}}
\newcommand{\es}{\end{split}}

\newcommand{\cH}{\mathcal{H}}
\newcommand{\cJ}{J}
\newcommand{\cG}{\mathcal{G}}
\newcommand{\cI}{I}
\newcommand{\cM}{\mathcal{M}}

\renewcommand{\vec}[1]{\bm{#1}}

\newcommand{\unit}[1]{\vec{\hat{ #1}}}

\DeclareMathOperator{\Tr}{Tr}
\newcommand{\up}{\uparrow}
\newcommand{\down}{\downarrow}
\newcommand{\llangle}{\left\langle}
\newcommand{\rrangle}{\right\rangle}
\newcommand{\jkp}{\hat J^{Kp}_\mu}

\newcommand{\jDp}{\hat J^{\Delta p}_\mu}

\newcommand{\AAA}{{\bf A}}
\newcommand{\RR}{{\bf R}}
\newcommand{\bb}{{\bf b}}
\newcommand{\qq}{{\bf q}}

\newcommand{\dsm}    	{D^{\rm(s)}_{\mu\nu}}
\newcommand{\ds}     	{$\dsm$\xspace}
\newcommand{\ceq}[1] 	{(\ref{#1})}
\newcommand{\rr}		{{\bf r}}

\begin{document}

\title{Superfluid Weight of Strongly Inhomogeneous Superconductors}

\author{Jonathan Schirmer}
\affiliation{Department of Physics, William \& Mary, Williamsburg, Virginia 23187, USA}
\author{Enrico Rossi}
\affiliation{Department of Physics, William \& Mary, Williamsburg, Virginia 23187, USA}

\begin{abstract}
In this work, we obtain the expression, within the linear response approximation, that allows the direct calculation of the superfluid weight for strongly inhomogeneous superconductors. Using this expression, we find that, in general, the correction to the superfluid weight due to the response of the superconductor's pairing potential to the perturbing vector potential is important in superconductors with a strongly inhomogeneous pairing potential. We consider two exemplary cases: the case when strong inhomogeneities in  the pairing potential are induced by a periodic  potential, and the case when superconducting vortices are induced by an external magnetic field. For both cases we show that the correction to the superfluid weight due to the response of the paring potential to the perturbing vector potential can be significant, it must be included to obtain quantitatively correct results, and that for the case when vortices are present the expression of the superfluid weight that does not include such correction returns qualitatively wrong results.
\end{abstract}
\graphicspath{{./Figures/}}

\maketitle

The Meissner effect is the hallmark signature of superconductivity.
It is described by the London equation
$\cJ_\mu = D_{\mu\nu}^{(s)}A_\nu$
where $\cJ_\mu$ is the $\mu$ component of the charge current, $A_{\nu}$
is the $\nu$ component of a static, transverse, long-wavelength, vector potential,
and $D_{\mu\nu}^{(s)}$ is the superfluid weight tensor. 
This equation shows that the superfluid weight quantifies the strength of the Meissner effect, and therefore
the ``robustness'' of the superconducting state.
As a consequence \ds can be seen as the key quantity that characterizes 
a superconductor~\cite{Scalapino1992,scalapino_insulator_1993}.
In two dimensions (2D) \ds is also the quantity
that fixes the critical temperature, $T_{BKT}$, at which the
Berezinskii-Kosterlitz-Thouless (BKT)~\cite{berezinsky1971,kosterlitz1973} transition, 
between superconducting and normal phase, takes place.
The essential role \ds plays in determining the crucial properties
of superconductors makes its correct and accurate determination very important.

For an isotropic superconductor with an isolated parabolic band crossing the Fermi
energy in the normal phase, at zero temperature, we have the conventional result ${\rm Tr}[D_{\mu\nu}^{(s)}]/d=e^2 n/m^*$,
where $d$ is the number of dimensions, $e$ is the electron's charge, $n$ is the electrons' density, 
and $m^*$ is the effective mass.
The realization of superconducting states in magic-angle twisted bilayer 
graphene~\cite{Cao2018,Lu2019,Yankowitz2019,Stepanov2020,Saito2020,Andrei2020,Park2021,Hao2021,Zhou2021,Cao2021,Zhou2022},
for which the bands are extremely flat so that $m^*\to\infty$,
has made clear the limitations of the conventional result.
In recent years, more general expressions for $D_{\mu\nu}^{(s)}$ accounting for the 
effect of quantum geometry in multi-band superconductors have 
been obtained~\cite{Peotta2015,Liang2017,wu2020c,bernevig2021h,Huhtinen2022,HerzogArbeitman2022,Lau2022,Toermae2022,torma_essay_2023,bouhon2023,tam_geometry-independent_2024,verma2024,kaplan2025a,yu2024c,jiang2025,penttila2025,yu2024b}.
These formulations show how, in flat-band systems like twisted bilayer graphene,
the contribution to $D_{\mu\nu}^{(s)}$ arising from the quantum geometry can be 
dominant~\cite{Hu2019a,Julku2020,xie2020,Verma2021,rossi2021l,hu2022x},
a result supported by recent experiments~\cite{tian2023,banerjee2025,tanaka2025}.

In many cases of interest, the pairing potential $\Delta$ cannot be assumed to be spatially uniform.
This is the case, for instance, when disorder is present~\cite{Ghosal1998,Paramekanti1998,Ghosal2001,Seibold2012,Lau2022, Datta2023},
or in the presence of superconducting vortices.
The presence of spatial inhomogeneities mixes the system's response to the longitudinal and transverse components of an external vector potential, a fact that makes the calculation of \ds more challenging \cite{Seibold2012}.
The reason is that whereas
the BCS mean-field treatment, within the linear response approximation, returns
the correct response of a superconducting system to a transverse vector field~\cite{schrieffer1964,Scalapino1992,scalapino_insulator_1993}, 
it is known that it returns an incorrect,  gauge-dependent, response
to a longitudinal vector field~\cite{schrieffer1964}.
This issue was addressed by several 
papers~\cite{Anderson1958,Anderson1958a,Rickayzen1959,Nambu1960,Kadanoff1961,Kulik1981,Emery1995,Zha1995,Arseev2006,wang2025}
that pointed out that for the general case, gauge invariance is restored by taking into account the vertex corrections for $\Delta$,
the so called collective-mode contributions,
i.e., by including the response of $\Delta$ to the vector field $\AAA$.
Later works considered the role of such contributions for specific 
cases~\cite{Guo2013,Yin2013,Boyack2016,Boyack2017,peotta_superconductivity_2022,Chan2022,tam_geometry-independent_2024,jiang2024,Nunchot2025}. 
In Ref.~\cite{Huhtinen2022} an expression for \ds was obtained 
that showed the importance of such contributions to restore the independence with respect to the position of the orbitals 
of the long-wavelength, zero frequency electromagnetic response of a superconductor. 
In this work, we obtain an expression of \ds that allows us to treat in a straightforward way 
also the challenging case when the phase of $\Delta$ varies rapidly in space, as is the case of superconductors with vortices.
To verify the accuracy of our expression we compare its predictions to the results obtained by calculating, via a numerical self-consistent approach, the free energy, $F$, as a function of the perturbing field ${\bf A}$ and then \ds as the second derivative of $F$ with respect to ${\bf A}$: $D_s^{\mu\nu}=(1/V)d^2 F/d A_\mu d A_\nu$, where $V$ is the system's volume. We do this for two important exemplary cases: the case when strong inhomogeneities in $\Delta$ are induced by a periodic superlattice potential applied to a 2D superconductor, and the case when superconducting vortices are induced by an external magnetic field perpendicular to a 2D superconductor. In both cases we find that the results obtained by numerically calculating the second derivative with respect to ${\bf A}$ of the free energy are the same as the ones obtained using the derived expression, but can be significantly different from the ones obtained using the expressions for \ds available in the literature that do not take into account the presence of inhomogeneities. For the case when vortices are present, the expression of \ds that does not include the response of $\Delta$ to ${\bf A}$ returns a qualitatively wrong result -- it erroneously predicts a finite value of \ds even for an infinite 2D array of unpinned vortices, in contrast to the expression that we present, that correctly returns $D_s^{\mu\nu}=0$ for this situation \cite{Teitel1983,Franz1995,jose_two-dimensional_2013}.

We describe the superconducting state using a Bogoliubov-de Gennes (BdG)
effective mean field Hamiltonian $\hat H_{\rm BdG}$. Given that the goal of this work is to obtain
the correct response of inhomogeneous superconductors to an external 
vector field, and not the identification of the many-body ground state, 
the use of the BdG approach is very pragmatic: it allows the modeling of
any superconducting state taking as inputs from experiments the values
of the parameters entering the model.
For concreteness, we consider a superconductor with s-wave pairing
originating from an on-site attractive interaction of strength $U>0$. 
For such a system
\begin{align} 
 \hat{H}_{\rm BdG} =& -\sum_{j l\sigma}t_j^{\ell}c^\dagger_{j+\beta_j^{\ell},\sigma}c_{j,\sigma}
                       -\sum_{j\sigma}\frac{1}{2}(\mu-V_j)c^\dagger_{j,\sigma} c_{j,\sigma} \nonumber  \\
                    &  -\sum_j \left[\Delta_j c^\dagger_{j,\uparrow}c^\dagger_{j,\downarrow} -\frac{|\Delta_j|^2}{2U}\right] + \text{H.c.}
 \label{mean field hamiltonian1}
\end{align}                     
where the subscript $j$ is shorthand for the position vector $\vec{r}_j$, $\sigma = \up, \down$ is the spin index,
$t_j^{\ell}$ is the hopping amplitude at position $\vec{r}_j$ along the bond $\vec{\beta}_j^{\ell}$,
$c^\dagger_{j,\sigma}$ ($c_{j,\sigma}$) is the creation (annihilation) operator
for an electron at position $\vec{r}_j$ with spin $\sigma$,
$\mu$ is the chemical potential, $V_j$ is an applied potential,
and $\Delta_j$ is the pairing potential at position $\vec{r}_j$ obtained
from the self-consistent equation $\Delta_j =U\langle c_{j,\downarrow}c_{j,\uparrow}\rangle$,
where the angle brackets denote equilibrium expectation values at temperature $T$.
In general, the lattice can have a basis and so $\vec{r}_j=\RR_i + \bb_m$,
where $\{\RR_i\}$ are the position vectors that identify the lattice and $\bb_m$ are the vectors 
for the positions of the basis elements within the primitive cell.
${\mathbf\beta}^{\ell}_j$ are bond vectors that connect sites within, and between, primitive cells, making the 
expression of $ \hat{H}_{\rm BdG}$, Eq.~\ceq{mean field hamiltonian1}, very general.
In the remainder, the primitive cell is chosen so that $\Delta(\rr_j)=\Delta(\rr_j + \RR_i)$, 
and therefore, when $\Delta$ is inhomogeneous, can be much larger than the crystal's primitive cell.

\ds relates the strength of the charge current $\hat{\vec{J}}$ to a static, transverse, vector field $\AAA$ with zero parallel momentum $\qq_\parallel$,
and perpendicular momentum $\qq_\perp\to 0$~\cite{Scalapino1992}: 
$\langle \hat J_\mu\rangle =D^{(s)}_{\mu\nu}A_\nu(\qq_\parallel=0,\qq_\perp\to 0,\omega=0)$,
where  $\langle \hat J_\mu\rangle$ is the expectation value of the $\mu$ component of the current,
and $\omega$ the frequency of the field $\AAA$.
\ds can therefore be obtained by calculating the linear response of $\hat{\vec{J}}$ 
to a transverse vector field $\AAA$.
For a tight-binding model, the presence of the field $\AAA$ can effectively be
taken into account by introducing a Peierls phase for
the hopping parameters: $t_j^{\ell}\to t_j^{\ell}e^{i\AAA\cdot{\vec{\beta}}^\ell_j}$.
In addition, it can induce a change in the pairing field that, as we will show, cannot be neglected for inhomogeneous superconductors.
Taking this into account, as was done in Ref. \cite{Huhtinen2022}, using $\hat J_\mu=-\delta \hat H_{BdG}/\delta A_\mu$, we find:
\begin{align}
 &\hat{\cJ}_\mu(\vec{r}_j)= \sum_{\ell, \sigma}\left( i (\vec{\beta}_j^\ell)_\mu~t_j^{\ell}e^{i \vec{A}(\vec{r}_j,t) \cdot \vec{\beta}_j^\ell} c^\dagger_{j+\beta_j^{\ell},\sigma} c_{j,\sigma} + H.c. \right) +\nonumber \\
  &\hspace{6mm}\sum_{j'} \left[\frac{\delta \Delta_{j'}}{\delta A_\mu (\vec{r}_j)} c^\dagger_{j',\uparrow}c^\dagger_{j',\downarrow}
  -\frac{1}{2U} \frac{\delta |\Delta_{j'}|^2}{\delta  A_\mu (\vec{r}_j)} + H.c.\right]
  \label{real space J-mt}
 \end{align}
To first order in $\AAA$ we have
\be
 \hat J_\mu = \jkp + \hat T^K_{\mu\nu}A^\nu + \jDp + \hat T^\Delta_{\mu\nu}A^\nu
\ee
where $\jkp$, $\hat T^K_{\mu\nu}A^\nu$ are the paramagnetic and diamagnetic currents, respectively, arising from the kinetic energy part of the BdG Hamiltonian,
and $\jDp$, $\hat T^\Delta_{\mu\nu}A^\nu$ the paramagnetic and diamagnetic currents due to the change of $\Delta_j$ induced by $\AAA$.
$\hat T^\Delta_{\mu\nu}A^\nu$ does not contribute to $\langle\hat J_\mu\rangle$
and to \ds (see SM) and so we can neglect it.
$\jDp$ is given by the second line of Eq.~\ceq{real space J-mt} by
evaluating the variational derivatives at $\AAA=0$.
$\jDp$ also does not contribute to $\langle\hat J_\mu\rangle$, 
but it does contribute to \ds. As we show below, its contribution to \ds
is critical when $\Delta$ is not homogeneous.

To obtain the current response to a vector field with vanishing momentum $\qq$,
it is convenient to express the current in momentum space. 
By performing the Fourier transform with respect to $\RR_i$ we can write%
\be
  c_{\RR_i+\bb_m,\sigma} = \frac{1}{\sqrt{N_c}}\sum_{\vec{k}} c_{m\sigma}(\vec{k})e^{i\vec{k}\cdot \bb_m}e^{i\vec{k}\cdot \RR_i}.
 \label{eq.embedding}
\ee
where $c_{m\sigma}(\vec{k})$ ($c^\dagger_{m\sigma}(\vec{k})$)
is the creation (annihilation) operator for an electron in the state $\ket{\vec{k}m\sigma}$
with momentum $\vec{k}$, orbital $m$, and spin $\sigma$.
The operator $c(\vec{k})_{m,\sigma}$ is defined apart from an overall phase factor.
In writing Eq.~\ceq{eq.embedding} we have chosen this overall phase factor to be $e^{i\vec{k}\cdot \bb_m}$,
given that this choice allows us
to write the full paramagnetic current operator, in momentum space, in the limit $\qq\to 0$, in the compact form:
\begin{equation}\label{nambu para current-mt}
    \hat{\cJ}_\mu^{p} (\vec{q}\to 0) = \frac{1}{N_c}\sum_{\vec{k}mm'}
    \psi_{m}^{\dagger}(\vec{k}) \cI_{mm'\mu}(\vec{k}) \psi_{m'}(\vec{k}) + C
\end{equation}
where $N_c$ is the number of unit cells,
$\psi_{m}^{\dagger}(\vec{k})=(c^\dagger_{m\uparrow}(\vec{k}), c^\dagger_{m\downarrow}(\vec{k}))$,
$C$ is a constant, and
\begin{equation}
  \cI_{mm';\mu}(\vec{k}) = 
  \begin{pmatrix} 
   \frac{\partial H_{m m'}}{\partial k_\mu}\big|_{\vec{k}} && 
   \frac{\delta \Delta_m}{\delta A_\mu}\big|_0\delta_{mm'} \vspace{2mm}  \\ 
   \frac{\delta \Delta_m^*}{\delta A_\mu}\big|_0\delta_{mm'} && 
   -\frac{\partial H_{mm'}^*}{\partial k_\mu}\big|_{-\vec{k}}
  \end{pmatrix}
  \label{eq.I}
\end{equation}
with $\{H_{mm'}(\vec{k})\}$ the matrix elements, in the basis $\left\{\ket{\vec{k}m\sigma}\right\}$,
of the normal state Hamiltonian, $\hat H$, and
$\delta \Delta_m/{\delta A_\mu}\big|_0\equiv
\delta \Delta_m/\delta A_\mu(\qq\to 0)|_{\AAA=0}$.

By setting the off-diagonal terms in $\cI_{mm';\mu}(\vec{k})$ equal to 0 we obtain the 
matrix $\cI^K_{mm';\mu}(\vec{k})$ that, when inserted in Eq.~\ceq{nambu para current-mt},
returns the expression of $\jkp$ in the limit $\qq\to 0$, $\omega=0$.
For the operator $\hat T^K_{\mu\nu}A^\nu$ we obtain
\begin{align} 
    \hat{T}_{\mu\nu}^K(\vec{q}) \!=\! & -\!\!\!\!\!\sum_{m, \ell, \sigma,\vec{k}}\!\!\!
    \frac{(\vec{\beta}_{m}^\ell)_\mu(\vec{\beta}_{m}^\ell)_\nu}{N_c} \nonumber
    \\&\times\left[ t_{m}^{\ell} e^{-i\vec{k}\cdot \vec{\beta}^\ell_{m}}c^\dagger_{m
    +\beta_{m}^\ell, \sigma}(\vec{k})c_{m \sigma}(\vec{k}+\vec{q})   \right.  \nonumber \\
    &+\left. t_{m}^{\ell*} e^{i(\vec{k}+\vec{q})\cdot \vec{\beta}^\ell_{m}} c^\dagger_{m \sigma}(\vec{k})  c_{m+
    \beta_{m}^\ell, \sigma}(\vec{k}+\vec{q})\right].
    \label{kinetic diamagnetic}
\end{align}

The superfluid weight is given by the sum of the paramagnetic current response $\Pi_{\mu\nu}(\qq\to0,\omega=0)$
and the expectation value of the operator $\hat T^K_{\mu\nu}$: 
\be
 \dsm = \Pi_{\mu\nu}(\qq\to0,\omega=0) + \langle \hat T^K_{\mu\nu}\rangle.
 \label{eq.Ds} 
\ee
We have
\begin{equation}
 \Pi_{\mu\nu}(\qq,0)= \frac{i}{N_c} \int_0^\infty dt
 \langle [\hat{\cJ}^{K,p}_\mu(\vec{q},t), \hat{\cJ}^p_\nu(\vec{-q},0)]  \rangle
 \label{eq.Pi}
\end{equation}
Notice that by replacing in Eq.~\ceq{eq.Pi} $\hat{\cJ}^p_\nu(\vec{-q},0)$ with $\hat{\cJ}^{K,p}_\nu(\vec{-q},0)$
we recover the expression of $\Pi$ that neglects the effect on $\Delta$ of $\AAA$.
In the limit $\qq\to 0$, $\omega=0$ we find: 
\begin{align} 
  &\Pi_{\mu\nu}(\qq\to0, 0) = \frac{1}{N_c}\sum_{\vec{k},ab} 
  \frac{n_F\left(E_a(\vec{k})\right)- n_F\left(E_b(\vec{k})\right)}{E_a(\vec{k}) - E_b(\vec{k})} \nonumber \\
  &\hspace{4mm}\bra{\phi_a(\vec{k})}  \cI^K_\mu(\vec{k}) \ket{\phi_b(\vec{k})}
   \bra{\phi_b(\vec{k})} \cI_\nu(\vec{k}) \ket{\phi_a(\vec{k})}
 \label{para response zero q-mt}
\end{align}
where $n_F$ is the Fermi-Dirac function, and $\{E_a\}$, $\{\ket{\phi_a(\vec{k})}\}$ are the eigenvalues, eigenvectors, respectively, of 
$\hat H_{BdG}$.
A key aspect of the expression for $\Pi$ given by Eq.~\ceq{para response zero q-mt} is that
the matrix $\cI_\nu(\vec{k})$, Eq.~\ceq{eq.I}, in Nambu space,
has non-zero off diagonal elements 
$\delta \Delta_m/{\delta A_\mu}\big|_0$.

To obtain the expression of $\delta \Delta_m/{\delta A_\mu}\big|_0$
we need to calculate the response of $\Delta$ to an external 
vector potential $\AAA$. We note that one can treat $\AAA$ as a parameter in the mean field calculation and use the finite difference approximation to determine $\delta \Delta_m/{\delta A_\mu}\big|_0$ (see the SM). This approach requires knowledge of the self-consistent solution for $\Delta$ at finite $\AAA$ as well as $\AAA=0$.
The linear response expression of $\delta \Delta_m/{\delta A_\mu}\big|_0$ requires only the solution at $\AAA=0$, and
is given by:
\begin{equation}
 \left.\frac{\delta\Delta_m}{\delta A_\mu}\right|_0=
 \frac{i}{N_c}\lim_{\vec{q}\to 0} \int_0^\infty dt
 \langle [\hat\Delta_{m}(\vec{q},t), \hat{\cJ}^p_\mu(\vec{-q},0)]  \rangle
 \label{eq.dD}
\end{equation}
where $\hat\Delta_m(\vec{q})=U\sum_{\vec{k}}c_{m\downarrow}(\vec{-k})c_{m\uparrow}(\vec{k}+\vec{q})$.
Notice that Eq.~\ceq{eq.dD} is equivalent to the inclusion of the anomalous components
of the vertex corrections~\cite{schrieffer1964,Raines2024}, the relevant components
for the mean-field treatment considered (see also the discussion on vertex corrections in the SM). 
The vertex corrections guarantee
that the expectation value of the full current operator
satisfies the Ward identities, and therefore charge conservation,
even when $\Delta$ is not homogeneous leading to mixing of responses to transverse and longitudinal vector fields.

Equation~\ceq{eq.dD} leads to a linear equation for $\delta \Delta_m/{\delta A_\mu}\big|_0$ of the form
\be 
\small
  K\!\!  
  \begin{pmatrix} 
   \delta \Delta_{\mu}^{(R)}  \vspace{2mm}  \\ 
   \delta \Delta_{\mu}^{(I)}  
  \end{pmatrix} \!\!=\! \begin{pmatrix}
      
   C_\mu^{(R)}\\ C_\mu^{(I)}
   \end{pmatrix};
  \hspace{2mm}
  K =
\begin{pmatrix} 
K_+^{(R)} &  & -K_-^{(I)}\\
K_+^{(I)} &  & K_-^{(R)}
\end{pmatrix}
 \label{eq.dD2} 
\ee
where $\delta\Delta^{(R)}$ and $\delta\Delta^{(I)}$ are the real and imaginary parts of the vector with components $\{\delta \Delta_m/{\delta A_\mu}\big|_0\}$, $C_\mu^{(R)}$ and $C_\mu^{(I)}$ are the real and imaginary parts of the vector with components 
\begin{align}
    &(C_\mu)_{m} = \frac{1}{N_c}\sum_{\vec{k},ab} \frac{n_F\left(E_a(\vec{k})\right)- n_F\left(E_b(\vec{k})\right)}{E_a(\vec{k}) - E_b (\vec{k})} 
 \nonumber \\
 &\hspace{10mm}\bra{\phi^{m}_a(\vec{k})} \tau_- \ket{\phi^{m}_b(\vec{k})}
 \bra{\phi_b(\vec{k})}  \cI^K_\mu(\vec{k}) \ket{\phi_a(\vec{k})}
 \label{eq.Ci}
\end{align}
and
$K_{\pm}^{(R/I)}=\mathcal{A}^{(R/I)} \pm \mathcal{B}^{(R/I)}$
with $\mathcal{A}^{(R/I)}$, $\mathcal{B}^{(R/I)}$ the real/imaginary parts of matrices with elements
\begin{align}
 &\mathcal{A}_{mm'} = \frac{-1}{N_c}\sum_{\vec{k},ab} \frac{n_F\left(E_a(\vec{k})\right)-n_F\left(E_b(\vec{k})\right)}{E_a(\vec{k})-E_b(\vec{k})} 
 \nonumber \\
 &\hspace{10mm}\bra{\phi^{m}_a(\vec{k})} \tau_- \ket{\phi^{m}_b(\vec{k})}
 \bra{\phi^{m'}_b(\vec{k})} \tau_+ \ket{\phi^{m'}_a(\vec{k})} - \frac{1}{U}\delta_{mm'}
 \label{eq.Aij}  \\ 
 &\mathcal{B}_{mm'} = \frac{-1}{N_c}\sum_{\vec{k},ab} \frac{n_F\left(E_a(\vec{k})\right)- n_F\left(E_b(\vec{k})\right)}{E_a(\vec{k}) - E_b(\vec{k})}
 \nonumber \\
 &\hspace{10mm}\bra{\phi^{m}_a(\vec{k})} \tau_- \ket{\phi^{m}_b(\vec{k})}
 \bra{\phi^{m'}_b(\vec{k})} \tau_- \ket{\phi^{m'}_a(\vec{k})}.
 \label{eq.Bij} 
\end{align}
In Eqs. \eqref{eq.Ci}-\eqref{eq.Bij}, $\ket{\phi^{m}_a(\vec{k})}$ is the component of $\ket{\phi_a(\vec{k})}$ on orbital $m$ and the $\tau$'s are Pauli matrices in particle-hole space. Na\"ively $\delta \Delta_m/{\delta A_\mu}\big|_0$
can be obtained by inverting Eq.~\ceq{eq.dD2}.
However, the square matrix $K$ is singular, it has rank one less than its dimension.
This reflects the fact that the vector ${\Delta_m}$, and therefore 
$\delta \Delta_m/{\delta A_\mu}\big|_0$
is defined apart from an overall phase $\alpha$ (see SM).
$\delta \Delta_m/{\delta A_\mu}\big|_0$, apart from the overall phase $\alpha$,
can be obtained by calculating the pseudoinverse of $K$ via
a singular value decomposition (see SM).

Equations~\ceq{eq.Ds}, 
\ceq{para response zero q-mt}, \ceq{eq.dD2}-\ceq{eq.Bij} 
allow the calculation of the full superfluid weight.
We can write 
$\dsm = D^{(s0)}_{\mu\nu}+\delta D^{s}_{\mu\nu}$, where 
$D^{(s0)}_{\mu\nu}$ is the value of \ds obtained neglecting the correction 
due to the response of $\Delta$ to $\AAA$, and
\be
    \delta D^{(s)}_{\mu\nu} = 2~\text{Re}\left[\sum_{m}(C_\mu)_{m}\left.\frac{\delta\Delta_m^*}{\delta A_\nu}\right|_0\right] 
    \label{eq.deltaDs}
\ee
is the correction due to the changes in the pairing potential induced by $\AAA$. 
We note that our result may be obtained using vertex corrections \cite{Nambu1960, schrieffer1964}, which we also discuss in the SM.
The correction $\delta D^{(s)}_{\mu\nu}$ given by Eq.~\ceq{eq.deltaDs} is gauge invariant (see SM).
To check that the inclusion of the correction $\delta D^{(s)}_{\mu\nu}$ 
returns the accurate quantitative value of \ds,
we compare the results obtained by
combining Eqs.~\ceq{eq.Ds}, 
\ceq{para response zero q-mt}, \ceq{eq.dD2}-\ceq{eq.Bij},
and the ones obtained using the relation $\dsm=(1/V)d^2 F/d A_\mu d A_\nu$.
The full derivative of $F$ with respect to $\AAA$ indicates that also the 
dependence of $F$ on $\AAA$ through $\Delta$ is included.
The derivatives $d^2 F/d A_\mu d A_\nu$ are calculated numerically for the ground state
that is obtained solving self-consistently the gap equation.

\begin{figure}
     \includegraphics[width=\linewidth]{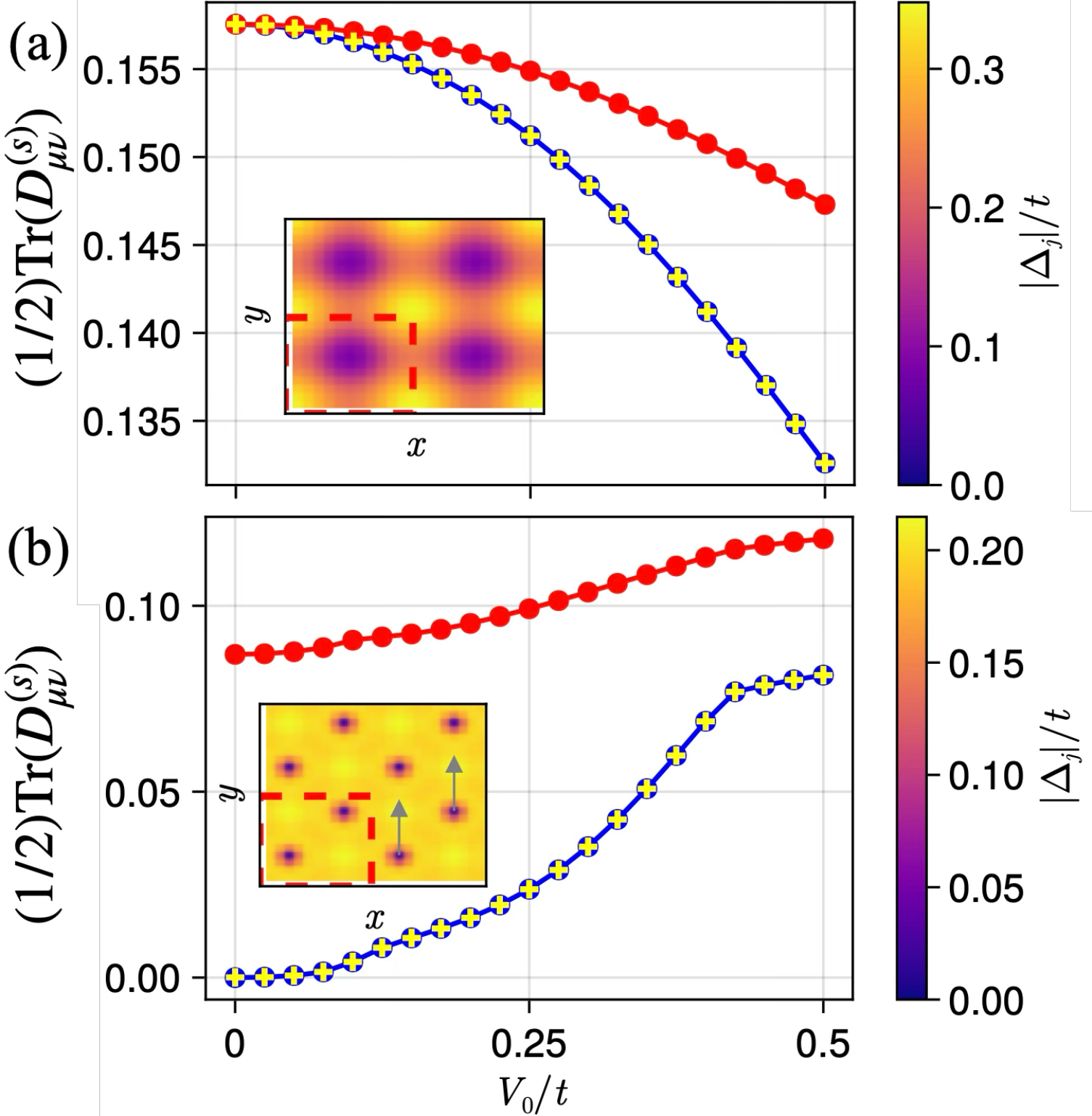}
     \caption{(a) $(1/2){\rm Tr}(\dsm)$ as a function of $V_0$ without (red points) 
     and with (blue points) the correction from the response of $\Delta$ for a 2D superconductor with
     ten $12 \times 12$ unit cells 
     in both the $x$ and $y$ directions ($N_c = 100$),
     $t = 1$, $\mu = -3.5$, and  $U = 3.4$.
     The yellow crosses show the values obtained by using the second derivative of the free energy. 
     Inset: color plot showing the spatial profile of $|\Delta_j|$ for $V_0 = 0.5$. The dashed red box represents one unit cell.
     (b) Same as (a) for the case of a superconductor with a vortex lattice induced
     a magnetic field $B_z= \Phi_0/144a^2$ ($\Phi_0=h/e$). 
     Inset: profile of $|\Delta_j|$ for $V_0 = 0$.
     The arrows in the inset show the direction of the vortices' motion when a perturbing vector potential $\AAA$ along the $x$ direction is applied.}
     \label{fig}
 \end{figure}

Figure~\ref{fig}~(a) shows the calculated values of $(1/2){\rm Tr}(\dsm)$ for a 2D superconductor
on a square lattice with lattice constant $a=1$ in the presence of the periodic potential
\begin{equation} \label{potential}
 V(\vec{r_j}) = -\frac{V_0}{2} \bigg[ \cos \Big(\frac{2\pi}{M}x_j \Big) + \cos \Big(\frac{2\pi}{M}y_j \Big) \bigg]
\end{equation}
with period $M$. In Eq.\ceq{potential} $\vec{r_j}=(x_j,y_j)$.
The potential has the effect of modulating the density of electrons, as well as the amplitude of the order parameter $\Delta_j$, thus rendering the superconductor inhomogeneous; see the inset of Fig. \ref{fig}~(a).
For $V_0=0$ the superconductor is homogenous, in this case $(C_\mu)_j=0$ (see SM) 
making $\delta \Delta_m/{\delta A_\mu}\big|_0=0$
and therefore $\dsm = D^{(s0)}_{\mu\nu}$. 
The results show that, indeed, for $V_0=0$ \ds, shown by the blue circles, 
coincides with $D^{(s0)}_{\mu\nu}$, shown by the red circles, and with the 
value obtained by calculating $d^2 F/d A_\mu d A_\nu$, yellow crosses.
However, for $V_0\ne 0$ $\Delta_j$ is inhomogeneous and so the correction $\delta D^{s}_{\mu\nu}$
is non negligible making $\dsm \neq D^{(s0)}_{\mu\nu}$, as shown in Fig.~\ref{fig}(a).
We see that for $V_0\ne 0$ only the value of \ds obtained by taking into account the corrections due to 
$\delta \Delta_m/{\delta A_\mu}\big|_0=0$
agrees with the value of \ds obtained by calculating
$d^2 F/d A_\mu d A_\nu$. 

The corrections to \ds due to $\delta \Delta_m/{\delta A_\mu}\big|_0$, i.e., $\jDp$, become  qualitatively very important
when vortices are present.
A lattice of unpinned vortices was found to have vanishing superfluid weight \cite{Teitel1983,Franz1995}.
This can be understood considering that for a vortex lattice in the $(x,y)$ plane induced by a background magnetic field $B_z$
in the direction perpendicular to the plane,
for a spatially, perturbing, constant, in-plane, vector potential $\AAA$, say along the $x$ direction, 
the vortices respond by shifting their position along the $y$ direction, as shown in the inset of Fig. \ref{fig}~(b), by an amount 
$ \Delta y = \frac{\hbar}{e} \frac{|\AAA|}{B_z}$.
Because this translation costs no free energy, given that it corresponds to the $\qq\to 0$ 
Goldstone mode associated to the translational symmetry spontaneously broken by the vortex lattice we have $\dsm=0$.

To study the superfluid weight in the presence of vortices, and a pinning periodic potential with period $M$ of the form 
given by Eq.~\ceq{potential}, we consider the case of a 2D superconductor on a square lattice in the presence of a perpendicular
background magnetic field $B_z$ \cite{Schirmer2024} (see SM for details). 
In Fig.~\ref{fig}~(b) we show the results for $D^{(s0)}_{\mu\nu}$,
\ds obtained taking into account the correction $\delta D^{s}_{\mu\nu}$,
and \ds obtained by taking the second derivative of $F$ with respect to $\AAA$.
The results show that for $V_0=0$, even though an unpinned vortex lattice is present,
$D^{(s0)}_{\mu\nu}$ is finite and quite large (red circles in the figure), of the same order as for the case
of a superconducting state with no vortices (see Fig.~\ref{fig}~(a)).
This contrasts with the expectation that superfluid weight should vanish.
The inclusion of the correction $\delta D^{(s)}_{\mu\nu}$ leads to 
$\dsm=0$ (blue circles), the correct value in the presence of an unpinned vortex lattice,
the same value that we find by calculating \ds as the second derivative of $F$ with respect to $\AAA$ (yellow crosses).
This is one of the key results of the present work: it shows that in the presence
of vortices the correction to \ds due to the response of $\Delta_j$ to $\AAA$
is essential to obtain the qualitatively correct value of the superfluid weight.
As $V_0$ increases, and the vortices start getting pinned, \ds also increases
from zero and starts getting closer to the value of $D^{(s0)}_{\mu\nu}$.
Notice, however, that even for $V_0=0.5 t$, the value of $D^{(s0)}_{\mu\nu}$ is still
about 60\% larger than the value given by the full expression of \ds, value that coincides
with the one obtained by calculating the second derivative of $F$ with respect to $\AAA$.

In summary, we have obtained an expression within linear response theory for the superfluid weight \ds designed for strongly inhomogeneous superconductors. We find that the corrections due to the response of the superconducting order parameter to the external vector potential \AAA play a significant role for superconducting states for which the superconducting pairing is inhomogeneous.
For the case of a superconducting vortex lattice with no pinning potential, we find that such corrections are essential to obtain the expected result of zero superfluid weight, showing the importance of such corrections for this experimentally very relevant case. For two-dimensional systems, the results presented show the importance of the response of the superconducting order parameter to the external  vector potential A when calculating  the critical temperature $T_{BKT}$ for the Berezinskii-Kosterlitz-Thouless phase transition of an inhomogeneous superconductor, given the connection between \ds and $T_{BKT}$.

This work was funded by the US Department of Energy, Office of Basic Energy Sciences, via Award DE-SC0022245.
The authors thank Kukka Huhtinen, Jainendra Jain, Chaoxing Liu, and Valerio Peri for very helpful discussions.


\bibliography{my.bib,additional.bib}



\clearpage
\newpage
\onecolumngrid
\section{Supplementary Material}

\tableofcontents

\section{Hamiltonian}
Let us suppose our system is defined on a lattice with sites $\{\vec{r}_j \}$, where the site/orbital labels are $j$, and with bonds described by a set of vectors $\{\vec{\beta}^{\ell}_j\}$ that start at site $j$, are labelled at each $j$ by the label $\ell$, and where $\text{Arg}(\vec{\beta}_j^{\ell}) \in (-\pi/2, \pi/2]$ in 2D. The latter condition is so that we consider ``forward hopping" only.  We take care of backward hopping, at the same time as ensuring Hermiticity, by adding the Hermitian conjugate of the forward hopping terms to the Hamiltonian. If two orbitals reside in the same location, one can unambiguously define $\vec{\beta}_j^{\ell}$ via point-splitting. We include an on-site superconducting pairing, originating from an on-site attractive interaction of magnitude $U>0$. The mean field Hamiltonian is
\be \label{mean field hamiltonian}
\hat{H}_{\rm MF} = -\sum_{j, \ell, \sigma}\left( t_j^{\ell}c^\dagger_{j+\beta_j^{\ell},\sigma} c_{j,\sigma} +\text{H.c.} \right) - \sum_{j, \sigma} \left(\mu - V_j\right)c^\dagger_{j,\sigma} c_{j,\sigma}   -\sum_{j} \Big(\Delta_j c^\dagger_{j,\uparrow}c^\dagger_{j,\downarrow}+\Delta_j^* c_{j,\downarrow}c_{j,\uparrow}-\frac{|\Delta_j|^2}{U} \Big)
\ee
where the subscript $j$ is shorthand for the position vector $\vec{r}_j$, the complex hopping amplitude along bond $\vec{\beta}_j^{\ell}$ is $t_j^{\ell}$, the chemical potential is $\mu$, $V_j$ is an applied potential, and, at zero temperature, 
$\Delta_j =U\langle c_{j,\downarrow}c_{j,\uparrow}\rangle$
with the angle bracket denoting equlibrium expectation values at temperature $T$.
These are the self-consistency equations. 

Suppose that the system possesses a translational invariance by vectors $\vec{a}_1,\vec{a}_2$ (specializing for 2D here). We may then define unit cells whose locations we specify by $\vec{R}_{i} = m\vec{a}_1+n\vec{a}_2$ with $m,n \in \mathbb{Z}$. The position of any orbital can then be given by
\begin{equation}
    \vec{r}_j = \vec{R}_{i}+\bb_m
\end{equation}
where $\bb_m$ 
specifies the location of the orbital \textit{within the unit cell.} The bonds and the hopping amplitudes then possess this translational invariance:
\beqr
 \vec{\beta}^\ell_{j+\vec{R}} &= \vec{\beta}^\ell_{j} \\ 
 t^\ell_{j+\vec{R}} &= t^\ell_{j} 
\eeqr
where the subscript $j+\vec{R}$ is shorthand for $\vec{r}_j+\vec{R}_{i}$. We will also suppose that the solution to the self-consistency equations possesses this same periodicity:
\begin{equation}
    \Delta_{j+\vec{R}} = \Delta_j
\end{equation}
We thus perform the Fourier transformation
\be \label{momentum space}
c_{j\sigma} = \frac{1}{\sqrt{N_c}} \sum_{\vec{k}} c_{m,\sigma}(\vec{k}) e^{i\vec{k}\cdot \vec{r}_j}
\ee
where $m$ is shorthand for $\vec{\tilde{r}}_j$. 
Using the basis of Nambu spinors
\begin{equation}
    \hat{\vec{\Psi}}(\vec{k}) = \begin{pmatrix}
        c_{1,\up}(\vec{k}) \\
        \vdots \\
        c_{N,\up}(\vec{k})\\
        c^\dagger_{1,\down}(-\vec{k})\\
        \vdots\\
        c^\dagger_{N,\down}(-\vec{k})
    \end{pmatrix}
\end{equation}
with $N$ the number of sites/orbitals per unit cell,
the Hamiltonian \eqref{mean field hamiltonian} can then be cast in Bogoliubov-de Gennes (BdG) form:
\begin{equation}
    \hat{\cH}_{\rm MF} = \sum_{\vec{k}} \hat{\vec{\Psi}}^\dagger(\vec{k}) H_{\rm BdG}(\vec{k})\hat{\vec{\Psi}}(\vec{k}) + \sum_{\vec{k}} \Tr[H(\vec{k})] + \frac{|\Delta_j|^2}{U}
\end{equation}
where
\begin{equation}
    H_{\rm BdG}(\vec{k}) = \begin{pmatrix}
        H(\vec{k}) && -\vec{\Delta} \\ -\vec{\Delta}^* && - H^*(-\vec{k})
    \end{pmatrix}
\end{equation}
is the BdG Hamiltonian, $H(\vec{k})$ is the normal state Hamiltonian, and
\begin{equation}
    \vec{\Delta} = \begin{pmatrix}
\Delta_1 & 0 & 0 & \cdots & 0 \\
0 & \Delta_2 & 0 & \cdots & 0 \\
0 & 0 & \Delta_3 & \cdots & 0 \\
\vdots & \vdots & \vdots & \ddots & \vdots \\
0 & 0 & 0 & \cdots & \Delta_N
\end{pmatrix}
\end{equation}

\section{Current Operator}
In this Supplementary Materials, $\vec{A}$ will refer to a probe electromagnetic potential. The orbital effect of any background magnetic field is accommodated in the model by the complex hopping amplitudes $t_j^{\ell}$. To determine the current, we must specify how the system couples to an external vector potential $\vec{A}(\vec{r},t)$. We suppose the hopping part of the Hamiltonian couples via a Peierls substitution
\be
t_j^{\ell} \rightarrow \exp \left(i \phi_j^\ell \right)t_j^{\ell}
\ee
where
\be \label{eq53}
 \phi_j^\ell = \int_{\vec{r}_j}^{\vec{r}_j + \vec{\beta}_j^\ell} \vec{A}(\vec{r},t) \cdot d\vec{r}
\ee
Notice that this change in the hopping amplitudes leads to a change in the groundstate $\ket{\rm G}$ which in general leads to a change in $\Delta_j$. 
We will compute this change in later sections.  We choose the path of integration to be the straight line connecting $\vec{r}_j$ to $\vec{r}_j + \vec{\beta}_j^\ell$ given by the vector $\vec{\beta}_j^\ell$. 

The system only couples to the average value of $\vec{A}(\vec{r},t)$ along the bonds. This is because
\be
\int_{\vec{r}_j}^{\vec{r}_j + \vec{\beta}_j^\ell} \vec{A}(\vec{r},t) \cdot d\vec{r} = \vec{A}_j^{\ell}(t) \cdot \vec{\beta}_j^\ell
\ee
where $\vec{A}_j^{\ell}(t)$ denotes the average value of $\vec{A}(\vec{r},t)$ along bond $\beta_j^{\ell}$. The current along bond $\beta_j^{\ell}$ (the current in lattice models is defined on the bonds) is given by

\begin{align}
\hat{\vec{\cJ}}_{j}^{\ell}&=-\frac{\delta \hat{\cH}_{\rm MF}}{\delta \vec{A}_j^{\ell}}= \sum_{\sigma}\left( i \vec{\beta}_j^\ell~t_j^{\ell}e^{i  \phi_j^\ell} c^\dagger_{j+\beta_j^{\ell},\sigma} c_{j,\sigma} +\text{H.c.} \right) 
+\sum_{i} \left(\frac{\delta \Delta_i}{\delta \vec{A}_j^{\ell}} c^\dagger_{i,\uparrow}c^\dagger_{i,\downarrow}+\frac{\delta\Delta_i^*}{\delta \vec{A}_j^{\ell}} c_{i,\downarrow}c_{i,\uparrow}-\frac{1}{U} \frac{\delta |\Delta_i|^2}{\delta \vec{A}_j^{\ell}} \right)
\\ \label{r current}
&=\vec{\beta}_j^\ell \sum_{\sigma}\left( i t_j^{\ell}e^{i  \phi_j^\ell} c^\dagger_{j+\beta_j^{\ell},\sigma} c_{j,\sigma} +\text{H.c.} \right)  
+\vec{\beta}_j^\ell\sum_{i} \left(\frac{\delta \Delta_i}{\delta \phi_j^{\ell}} c^\dagger_{i,\uparrow}c^\dagger_{i,\downarrow}+\frac{\delta\Delta_i^*}{\delta \phi_j^{\ell}} c_{i,\downarrow}c_{i,\uparrow}-\frac{1}{U} \frac{\delta |\Delta_i|^2}{\delta \phi_j^{\ell}} \right) \\
&\equiv \hat{\vec{\cJ}}_{j}^{\ell K} + \hat{\vec{\cJ}}_{j}^{\ell\Delta}
\end{align}
The first sum in \eqref{r current} is denoted $\hat{\vec{\cJ}}_{j}^{\ell K}$ and is the contribution of the current coming from the kinetic energy. The second sum, which is the contribution of the current from the pairing potential, is denoted $\hat{\vec{\cJ}}_{j}^{\ell\Delta}$. The distinction between these two contributions, which we may call \textit{kinetic current} and \textit{pairing current}, respectively, will play an important role in this SM. We see explicitly that the current is directed along the bonds. We have also taken the chemical potential to be fixed. 

We make a few additional comments on \eqref{r current}:
 \begin{enumerate} 
 \item In general, ${\delta \Delta_j}/{\delta \phi_{j'}^{\ell}}\neq 0$ for $j \neq j'$. For example, when varying $\phi_j^{\ell}$, we should at the very least expect the pairing potential $\Delta_i$ to respond at the starting and ending sites of the bond $\beta_j^{\ell}$. 
\item The self-consistency equations imply that the average pairing current vanishes $\llangle \hat{\vec{\cJ}}_j^{\ell \Delta}\rrangle \equiv 0$. Thus the average current is given by the average kinetic current 
\be
\llangle \hat{\vec{\cJ}}_j^{\ell}\rrangle = \vec{\beta}_j^\ell \sum_{\sigma}\left( i t_j^{\ell}e^{i  \phi_j^\ell} \Big\langle c^\dagger_{j+\beta_j^{\ell},\sigma} c_{j,\sigma} \Big\rangle +\text{c.c.} \right)  = \llangle \hat{\vec{\cJ}}_j^{\ell K}\rrangle
\ee
However, the pairing current does contribute to current correlations in the system, as we shall see. 
\end{enumerate}

If we are only interested in the responses of the system to vector potentials which vary slowly in space compared to the length of the bonds, we may replace $\vec{A}_j^{\ell} \rightarrow \vec{A}(\vec{r}_j,t)$, up to negligible error of $\mathcal{O}\big(|\vec{\beta}_j^\ell|/\lambda\big)$, where $\lambda$ is the characteristic length scale of variations in $\vec{A}(\vec{r},t)$. We can then meaningfully define the current at the lattice sites by
\be \label{real space J}
\begin{split}
 \hat{\cJ}_\mu(\vec{r}_j)=-\frac{\delta \hat{\cH}_{\text{MF}}}{\delta A_\mu(\vec{r}_j)}  = 
 &\sum_{\ell, \sigma}\left( i (\vec{\beta}_j^\ell)_\mu~t_j^{\ell}e^{i \vec{A}(\vec{r}_j,t) \cdot \vec{\beta}_j^\ell} c^\dagger_{j+\beta_j^{\ell},\sigma} c_{j,\sigma} +\text{H.c.} \right) \\
 &+ \sum_{j'} \left(\frac{\delta \Delta_{j'}}{\delta A_\mu (\vec{r}_j)} c^\dagger_{i,\uparrow}c^\dagger_{i,\downarrow}
  +\frac{\delta\Delta_{j'}^*}{\delta  A_\mu (\vec{r}_j)} c_{i,\downarrow}c_{i,\uparrow}-\frac{1}{U} \frac{\delta |\Delta_{j'}|^2}{\delta  A_\mu (\vec{r}_j)} \right) \\
 &\equiv \hat{\cJ}_{\mu}^{K}(\vec{r}_j) + \hat{\cJ}_{\mu}^{\Delta}(\vec{r}_j)
 \end{split}
\ee 
If the system is perturbed by a vector potential of the form $A_{\mu}(\vec{r},t) = A_\mu(\vec{q}, t)e^{-i\vec{q}\cdot\vec{r} }$ then the corresponding coupling is 
\be
\sum_{j,\mu}  \frac{\delta \hat{\cH}_{\text{MF}}}{\delta A_\mu(\vec{r}_j, t)}A_\mu(\vec{q}, t)e^{-i\vec{q}\cdot\vec{r}_j} = -\sum_{\mu}A_\mu(\vec{q}, t)\sum_{j}  \hat{\cJ}_\mu(\vec{r}_j, t)e^{-i\vec{q}\cdot\vec{r}_j} = -\sum_{\mu}A_\mu(\vec{q}, t) \hat{\cJ}_\mu (\vec{q}, t)
\ee
where $\mu$ is a spatial index ($\mu = x,y,\ldots$). Thus
\be
\begin{split} \label{q current}
\hat{\cJ}_\mu (\vec{q}) = &\sum_{j, \ell, \sigma}\left( i (\vec{\beta}_j^\ell)_\mu~t_j^{\ell}e^{i \vec{A}(\vec{r}_j,t) \cdot \vec{\beta}_j^\ell} c^\dagger_{j+\beta_j^{\ell},\sigma} c_{j,\sigma} +\text{H.c.} \right)e^{-i\vec{q}\cdot\vec{r}_j} \\
 &+ \sum_{j,j'} \left(\frac{\delta \Delta_{j'}}{\delta A_\mu (\vec{r}_j)} c^\dagger_{i,\uparrow}c^\dagger_{i,\downarrow}+
    \frac{\delta\Delta_{j'}^*}{\delta  A_\mu   (\vec{r}_j)} c_{i,\downarrow}c_{i,\uparrow}-\frac{1}{U} \frac{\delta |\Delta_{j'}|^2}{\delta  A_\mu (\vec{r}_j)} \right)e^{-i\vec{q}\cdot\vec{r}_j} \\
 &\equiv \hat{\cJ}^K_\mu (\vec{q})+\hat{\cJ}^\Delta_\mu (\vec{q})
 \end{split}
\ee

%
\subsection{Paramagnetic Current}
It is useful to decompose the current into paramagnetic and diamagnetic components
\begin{equation}
    \hat{\cJ}_\mu (\vec{q}) = \hat{\cJ}^p_\mu (\vec{q}) + \hat{T}_{\mu\nu}(\vec{q})A_{\nu}(\vec{q})
\end{equation}
where there is an implied sum over $\nu$. Both $\hat{\cJ}^p_\mu (\vec{q})$ and $\hat{T}_{\mu\nu}(\vec{q})$ admit further decomposition into kinetic and pairing contributions, as in the previous section. The kinetic component of the paramagnetic current is
\begin{align} \label{real current}
 \hat{\cJ}^{K,p}_\mu (\vec{q}) &= \sum_{j, \ell, \sigma}\left( i (\vec{\beta}_j^\ell)_\mu~t_j^{\ell}c^\dagger_{j+\beta_j^{\ell},\sigma} c_{j,\sigma} - i (\vec{\beta}_j^\ell)_\mu~t_j^{\ell *} c^\dagger_{j,\sigma}c_{j+\beta_j^{\ell},\sigma} \right) e^{-i\vec{q}\cdot\vec{r}_j} \\
 &=\frac{i}{{N_c}} \sum_{m, \ell, \sigma} \sum_{\vec{k}}(\vec{\beta}_{m}^\ell)_\mu \left[ t_{m}^{\ell} e^{-i\vec{k}\cdot \vec{\beta}^\ell_{m}}c^\dagger_{m+\beta_{m}^\ell \sigma}(\vec{k})c_{m \sigma}(\vec{k}+\vec{q})  -t_{m}^{\ell*} e^{i(\vec{k}+\vec{q})\cdot \vec{\beta}^\ell_{m}} c^\dagger_{m \sigma}(\vec{k})  c_{m+\beta_{m}^\ell \sigma}(\vec{k}+\vec{q})\right] 
\end{align}
where we have used \eqref{momentum space}.

The pairing contribution to the paramagnetic current is
\be
\begin{split}
 \hat{\cJ}^{\Delta,p}_\mu (\vec{q})&=\sum_{j'} \left[ \left( \sum_j \frac{\delta \Delta_{j'}}{\delta A_\mu (\vec{r}_j)}\Bigg|_{A=0} e^{-i\vec{q}\cdot\vec{r}_j} \right) c^\dagger_{j',\uparrow}c^\dagger_{j',\downarrow}+ \left(\sum_j \frac{\delta\Delta_{j'}^*}{\delta  A_\mu (\vec{r}_j)}\Bigg|_{A=0}e^{-i\vec{q}\cdot\vec{r}_j}\right) c_{j',\downarrow}c_{j',\uparrow}-\left(\sum_j \frac{1}{U} \frac{\delta |\Delta_{j'}|^2}{\delta  A_\mu (\vec{r}_j)}\Bigg|_{A=0} e^{-i\vec{q}\cdot\vec{r}_j} \right) \right] \\
&= \sum_{j'} \left[  \frac{\delta \Delta_{j'}}{\delta A_\mu (\vec{q})}\Bigg|_{A=0} c^\dagger_{j',\uparrow}c^\dagger_{j',\downarrow}+ \frac{\delta\Delta_{j'}^*}{\delta  A_\mu (\vec{q})}\Bigg|_{A=0}c_{j',\downarrow}c_{j',\uparrow}- \frac{1}{U} \frac{\delta |\Delta_{j'}|^2}{\delta  A_\mu (\vec{q})}\Bigg|_{A=0}  \right]  
\end{split}
\ee
where the derivatives with respect to $\vec{A}$ are evaluated at $\vec{A}=0$. Hereafter, all derivatives with respect to $\vec{A}$ will be evaluated at $\vec{A}=0$; thus, we omit the evaluation symbol on the derivatives from now on. We have used that the variation $\delta \Delta_{j'}$ with respect to a vector potential of the form $\delta A_{\mu}(\vec{r},t) = \delta A_\mu(\vec{q}, t)e^{-i\vec{q}\cdot\vec{r} }$ is
\be
\delta \Delta_{j'} = \sum_{j'} \frac{\delta \Delta_{j'}}{\delta A_{\mu}(\vec{r}_j)} \delta A_\mu(\vec{q}, t)e^{-i\vec{q}\cdot\vec{r}_j }
\ee
so that
\be
\sum_j \frac{\delta \Delta_{j'}}{\delta A_\mu (\vec{r}_j)} e^{-i\vec{q}\cdot\vec{r}_j} = \frac{\delta \Delta_{j'}}{\delta A_\mu (\vec{q})}
\ee
Recall that we have assumed that $\Delta_j$ at $\vec{A}=0$ is periodic with the same periodicity as the hopping amplitudes $ t^\ell_{j} $, i.e. we can still define a unit cell by the vectors $\vec{a}_1,\vec{a}_2$ (in 2D) in the presence of pairing. 
More generally we take $\Delta_i$ to be periodic, and take its period to be commensurate with that of the hopping amplitudes. This situation requires a modification of $\vec{a}_1,\vec{a}_2$, but otherwise no generality is lost.
Thus $\Delta_j$ is only a function of the intra-unit cell label. In other words,
\begin{equation}
    \Delta_{j} = \Delta_{m}
\end{equation}
From this, it follows that $\frac{\delta \Delta_{j'}}{\delta A_\mu (\vec{q})}$ can be written as a periodic function times a plane wave:

\begin{equation}
\frac{\delta \Delta_{j'}}{\delta A_\mu (\vec{q})} = \frac{\delta \Delta_{m}}{\delta A_\mu (\vec{q})}e^{-i\vec{q}\cdot\RR_i} 
\end{equation}
Thus the pairing contribution to the current, after using \eqref{momentum space}, is
\begin{equation} \label{pairing current}
    \hat{\cJ}^{\Delta,p}_\mu (\vec{q})=\frac{1}{N_c}\sum_{m, \vec{k}} \left[  \frac{\delta \Delta_{m}}{\delta A_\mu (\vec{q})}c^\dagger_{m,\uparrow}(\vec{k})c^\dagger_{m,\downarrow}(-\vec{k}-\vec{q})+ \frac{\delta\Delta_{m}^*}{\delta  A_\mu (\vec{q})}c_{m,\downarrow}(-\vec{k})c_{m,\uparrow}(\vec{k}+\vec{q}) \right]- \sum_{j} \frac{1}{U} \frac{\delta |\Delta_j|^2}{\delta  A_\mu (\vec{q})}
\end{equation}

The paramagnetic current is then
\begin{multline} \label{para current}
\hat{\cJ}_\mu^{\text{p}} (\vec{q}) = \frac{i}{N_c} \sum_{m, \ell, \sigma} \sum_{\vec{k}}(\vec{\beta}_{m}^\ell)_\mu \left[ t_{m}^{\ell} e^{-i\vec{k}\cdot \vec{\beta}^\ell_{m}}c^\dagger_{m+\beta_{m}^\ell \sigma}(\vec{k})c_{m \sigma}(\vec{k}+\vec{q})  -t_{m}^{\ell*} e^{i(\vec{k}+\vec{q})\cdot \vec{\beta}^\ell_{m}} c^\dagger_{m \sigma}(\vec{k})  c_{m+\beta_{m}^\ell \sigma}(\vec{k}+\vec{q})\right]
 \\+\frac{1}{N_c}\sum_{m, \vec{k}} \left[  \frac{\delta \Delta_{m}}{\delta A_\mu (\vec{q})} c^\dagger_{m,\uparrow}(\vec{k})c^\dagger_{m,\downarrow}(-\vec{k}-\vec{q})+ \frac{\delta\Delta_{m}^*}{\delta  A_\mu (\vec{q})}c_{m,\downarrow}(-\vec{k})c_{m,\uparrow}(\vec{k}+\vec{q}) \right]- \sum_{i} \frac{1}{U} \frac{\delta |\Delta_i|^2}{\delta  A_\mu (\vec{q})}
\end{multline} 
It is convenient to write the kinetic part of the current operator in the following way (what follows is merely a matter of convenient notation)
\begin{equation} 
\begin{split}
\hat{\cJ}^{K,p}_\mu (\vec{q})=& \frac{1}{N_c}\sum_{\vec{k}, \sigma} \sum_{m m'}J_{m m';\mu}(\vec{k}, \vec{q}) c^\dagger_{m \sigma}(\vec{k})c_{m'\sigma}(\vec{k}+\vec{q}) \\
=& \frac{1}{N_c}\sum_{\vec{k}} \sum_{m m'}\left[J_{m m';\mu}(\vec{k}, \vec{q}) c^\dagger_{m \up}(\vec{k})c_{m'\up}(\vec{k}+\vec{q})- J_{m' m;\mu}(-\vec{k}-\vec{q}, \vec{q}) c_{m \down}(-\vec{k})c^\dagger_{m'\down}(-\vec{k}-\vec{q})\right] + \text{c-numbers}
\end{split}
\end{equation}
where $m$ and $m'$ are site/orbital labels. The matrix elements $J_{m m';\mu}(\vec{k}, \vec{q})$ can be read off from \eqref{para current}. Now, Hermiticity of the current operator implies $J_{m' m;\mu}(-\vec{k}-\vec{q}, \vec{q})= J_{m m';\mu}(-\vec{k}, -\vec{q})^*$.  Thus the total paramagnetic current can be written in the basis $$ \psi_{m}(\vec{k}) = \begin{pmatrix} c_{m\up}(\vec{k}) \\ c^\dagger_{m\down}(-\vec{k})\end{pmatrix} $$
as
\begin{equation}\label{nambu para current}
\begin{split}
    \hat{\cJ}_\mu^{p} (\vec{q}) &= \frac{1}{N_c}\sum_{\vec{k}}\sum_{mm'}\psi^{\dagger}_{m}(\vec{k})\begin{pmatrix} J_{mm';\mu}(\vec{k},\vec{q}) && \frac{\delta \Delta_{m}}{\delta A_\mu (\vec{q})} \delta_{mm'}  \\ \frac{\delta \Delta^*_{m}}{\delta A_\mu (\vec{q})} \delta_{mm'} && -J^*_{mm';\mu}(-\vec{k},-\vec{q})  \end{pmatrix}\psi_{m'}(\vec{k}+\vec{q}) + \text{c-numbers} \\
   &\equiv \frac{1}{N_c} \sum_{\vec{k}}\sum_{mm'}\psi_m^{\dagger}(\vec{k}) \cI_{mm';\mu}(\vec{k}, \vec{q}) \psi_{m'}(\vec{k}+\vec{q})
   + \text{c-numbers}
    \end{split}
\end{equation}
where 
\begin{equation} 
    \cI_{m m';\mu}(\vec{k}, \vec{q}) = \begin{pmatrix} J_{mm';\mu}(\vec{k},\vec{q}) && \frac{\delta \Delta_{m}}{\delta A_\mu (\vec{q})} \delta_{mm'}  \\ \frac{\delta \Delta^*_{m}}{\delta A_\mu (\vec{q})} \delta_{mm'} && -J^*_{mm';\mu}(-\vec{k},-\vec{q})  \end{pmatrix}
\end{equation}
It is convenient to do the same for the kinetic current only:
\begin{equation}
     \hat{\cJ}_\mu^{K,p} (\vec{q}) = \frac{1}{N_c} \sum_{\vec{k}}\sum_{mm'}\psi_m^{\dagger}(\vec{k}) \cI^K_{mm';\mu}(\vec{k}, \vec{q}) \psi_m'(\vec{k}+\vec{q})
   + \text{c-numbers}
\end{equation}
where 
\begin{equation} 
    \cI^K_{m m';\mu}(\vec{k}, \vec{q}) = \begin{pmatrix} J_{mm';\mu}(\vec{k},\vec{q}) && 0 \\ 0 && -J^*_{mm';\mu}(-\vec{k},-\vec{q})  \end{pmatrix}
\end{equation}
%

\subsection{Diamagnetic Current}
The diamagnetic current should also be computed in order to calculate the linear response to an electromagnetic field. The diamagnetic current is obtained by carrying out the expansion in $\vec{A}$ to linear order in \eqref{q current} and going over to momentum space. The result for the prefactor of $A_\nu(\vec{q})$ coming from the kinetic current in such an expansion is
\begin{equation} \label{kinetic doamagnetic}
    \hat{T}_{\mu\nu}^K(\vec{q}) = -\frac{1}{{N_c}} \sum_{m, \ell, \sigma} \sum_{\vec{k}}(\vec{\beta}_{m}^\ell)_\mu(\vec{\beta}_{m}^\ell)_\nu\left[ t_{m}^{\ell} e^{-i\vec{k}\cdot \vec{\beta}^\ell_{m}}c^\dagger_{m+\beta_{m}^\ell \sigma}(\vec{k})c_{m \sigma}(\vec{k}+\vec{q})  + t_{m}^{\ell*} e^{i(\vec{k}+\vec{q})\cdot \vec{\beta}^\ell_{m}} c^\dagger_{m \sigma}(\vec{k})  c_{m+\beta_{m}^\ell \sigma}(\vec{k}+\vec{q})\right] 
\end{equation}

There is also a pairing contribution $\hat{T}_{\mu\nu}^{\Delta}(\vec{q})$. However, since $\langle \hat{\cJ}^\Delta_\mu (\vec{q})\rangle \equiv 0$, this term does not contribute to the response of the current, and thus is irrelevant for our calculation of the superfluid weight.

\section{Superfluid Weight}
The superfluid weight may be computed by calculating the response of the current $\langle \hat{\cJ}_{\mu}\rangle \equiv \langle \hat{\cJ}^K_{\mu}\rangle$ to a static vector potential ($\omega = 0$) in the long wavelength limit ($\vec{q}\rightarrow 0$) \cite{Scalapino1992,scalapino_insulator_1993}
\begin{equation}
    \delta\langle \hat{\cJ}_{\mu}(\vec{q}\rightarrow 0,\omega=0)\rangle\equiv\delta\langle \hat{\cJ}^K_{\mu}(\vec{q}\rightarrow 0,\omega=0)\rangle = 
    \dsm A_\nu(\vec{q}= 0,\omega=0)
\end{equation}
where
\begin{equation} \label{K definition}
    \dsm = \llangle \hat{T}_{\mu\nu}^{K}(\vec{q}=0,\omega=0)\rrangle + \lim_{\vec{q} \rightarrow 0} \Pi_{\mu\nu}(\vec{q}, \omega=0)
\end{equation}
We will see that the pairing potential modifies the paramagnetic current-current correlation function $\Pi_{\mu\nu}(\vec{q}, \omega)$.

%
\subsection{Paramagnetic Current-Current Correlation Function}
We will compute the paramagnetic current-current correlation function while taking into account the dependence of the pairing potential on $A_{\mu}(\vec{q})$. 
We compute the following response function in imaginary time $\tau$:
\begin{equation}
\Pi_{\mu\nu}(\vec{q}, \tau)=-\frac{1}{N_c}\llangle \mathcal{T}_\tau \hat{\cJ}^{K,p}_\mu(\vec{q},\tau) \hat{\cJ}^p_\nu(\vec{-q},0)  \rrangle
\end{equation}
The time-ordering symbol in imaginary time is $\mathcal{T}_\tau$. 
Note that the first factor in the expectation value above is the kinetic current operator 
$\hat{\cJ}^{K,p}_\mu$, since we are computing the response of 
$\langle \hat{\cJ}^K_{\mu}\rangle$,  
whereas the second factor is the full current 
$\hat{\cJ}^{p}_\mu = \hat{\cJ}^{K,p}_\mu +\hat{\cJ}^{\Delta,p}_\mu$, 
since the vector potential couples to the full current. 

The response function may be represented in the Matsubara frequency domain, using \eqref{nambu para current}, as
\begin{equation}
 \label{pi with G}
\Pi_{\mu\nu}(\vec{q}, i\omega_n) = \frac{1}{\beta N_c} \sum_{mm'll'} \sum_{\vec{k}, ik_{n'}} \Tr \bigg[ \cG_{m' l'}(\vec{k}, ik_{n'}) \cdot  \cI^K_{l' l ;\mu}(\vec{k}, \vec{q}) \cdot \cG_{lm}(\vec{k}+\vec{q}, ik_{n'} + i\omega_n) \cdot \cI_{m m';\nu}(\vec{k}+\vec{q}, -\vec{q})\bigg]
\end{equation}
The centered dots indicate matrix multiplication in particle/hole space and $\cG$ can be expressed in the basis
\begin{equation}
 \psi_{m}(\vec{k}) = \begin{pmatrix} c_{m\up}(\vec{k}) \\ c^\dagger_{m\down}(-\vec{k})\end{pmatrix} \end{equation}
as
\begin{equation}
    \cG_{mm'}(\vec{k}, ik_{n'}) = - \int_0^\beta d\tau \llangle \mathcal{T}_\tau \psi_m(\vec{k},\tau)  \psi^\dagger_{m'}(\vec{k},0)\rrangle e^{ik_{n'} \tau}
\end{equation}
 which is a $2 \times 2$ matrix-valued Green's function; $\alpha$ and $\beta$ label the orbital/site within the unit cell. $\omega_n = 2\pi n / \beta$ is a bosonic frequency and $k_{n'} = (2n'+1)\pi/\beta$ is a fermionic frequency. It is convenient to express $\cG$ in terms of eigenstates 
 $\ket{\phi_a(\vec{k})}$ with particle and hole components at site/orbital $m$
 \be
 \ket{\phi_a^m(\vec{k})} = \begin{pmatrix} u^m_a(\vec{k}) \\ v^m_a(\vec{k})\end{pmatrix}
 \ee
 and energies $E_a(\vec{k})$ of the BdG Hamiltonian as
\begin{equation}
    \cG_{m m'}(\vec{k}, ik_m) = \sum_a \frac{\ket{\phi^m_a(\vec{k})}\bra{\phi_a^{m'}(\vec{k})}}{ik_{n'} - E_a(\vec{k})}
\end{equation}
Inserting this into \eqref{pi with G} and doing the sum over $k_{n'}$, we obtain, in the static limit $\omega_n = 0$,
\begin{align} \label{para response}
\Pi_{\mu\nu}(\vec{q}, i\omega_n = 0)=\frac{1}{N_c}\sum_{\vec{k}}\sum_{ab} \frac{n_F\left(E_a(\vec{k})\right)- n_F\left(E_b(\vec{k}+\vec{q})\right)}{E_a(\vec{k}) - E_b(\vec{k}+\vec{q})} \bra{\phi_a(\vec{k})}  \cI^K_{\mu}(\vec{k}, \vec{q}) \ket{\phi_b(\vec{k}+\vec{q})}
\\\times\bra{\phi_b(\vec{k}+\vec{q})} \cI_{\nu}(\vec{k}+\vec{q}, -\vec{q}) \ket{\phi_a(\vec{k})} \nonumber
\end{align}
Where we have introduced the shorthand notation 
$\bra{\phi_b(\vec{k})} M \ket{\phi_a(\vec{k}')} = \sum_{mm'}\bra{\phi^{m}_b(\vec{k})} M_{mm'} \ket{\phi^{m'}_a(\vec{k}')} $
We then take the limit $\vec{q} \rightarrow 0$, where $\cI_{m\beta;\mu}(\vec{k}, \vec{q}\rightarrow 0)$ may be expressed as
\begin{equation}
  \cI_{mm';\mu}(\vec{k}, 0) = \begin{pmatrix} \frac{\partial H_{mm'}}{\partial k_\mu}\big|_{\vec{k}} && \frac{\delta \Delta_m}{\delta Q_\mu}\delta_{mm'} \vspace{2mm}  \\ \frac{\delta \Delta_m^*}{\delta Q_\mu}\delta_{mm'} && -\frac{\partial H_{mm'}^*}{\partial k_\mu}\big|_{-\vec{k}}\end{pmatrix}
\end{equation}
where $H$ is the normal state Hamiltonian. We have called $A_\mu (\vec{q} = 0) = Q_\mu$. Note that $\frac{\delta \Delta_m}{\delta Q_\mu}\equiv\left.\frac{\delta\Delta_m}{\delta A_\mu}\right|_0$ from the main text. We have

\begin{equation} \label{para response zero q}
\Pi_{\mu\nu}(\vec{q}\rightarrow 0, i\omega_n = 0)=\frac{1}{N_c}\sum_{\vec{k}}\sum_{ab} \frac{n_F\left(E_a(\vec{k})\right)- n_F\left(E_b(\vec{k})\right)}{E_a(\vec{k}) - E_b(\vec{k})} \bra{\phi_a(\vec{k})}  \cI^K_\mu(\vec{k}, 0) \ket{\phi_b(\vec{k})}
\bra{\phi_b(\vec{k})} \cI_\nu(\vec{k}, 0) \ket{\phi_a(\vec{k})}
\end{equation}
Note that the limit $\vec{q}\rightarrow 0$ implies that $\frac{n_F\left(E_a(\vec{k})\right)- n_F\left(E_b(\vec{k})\right)}{E_a(\vec{k}) - E_b(\vec{k})}$ should be understood as $\frac{\partial n_F}{\partial E}$ whenever $E_b(\vec{k}) = E_a(\vec{k})$.

%
\subsection{Calculating the Response of $\Delta$}
In order to complete the calculation, we need to compute $\frac{\delta \Delta_m}{\delta Q_\mu}$. This can be done in a number of ways including finite difference as mentioned in the main text (also see below), differentiating the gap equations directly, solving the Bethe-Salpeter equation for the vertex correction, or using linear response. Here, we use linear response. The operator $\hat{\Delta}_{m}$ can be written in Nambu form by writing
\begin{equation}
    \hat{\Delta}_{m} (\vec{q}) = U \sum_{\vec{k}}c_{m,\downarrow}(-\vec{k})c_{m,\uparrow}(\vec{k}+\vec{q}) = 
    U\sum_{\vec{k}}\sum_{ll'} \psi^\dagger_{l}(\vec{k}) \cdot \mathcal{M}^{m}_{ll'} \cdot \psi_{l'}(\vec{k}+\vec{q})
\end{equation}
where $\mathcal{M}^{m}_{ll'} = \delta_{ll'}\delta_{l,m}\tau_-$ and $\tau_-= \frac{1}{2} (\tau_x - i\tau_y)$ where $\tau_x$ and $\tau_y$ are the Pauli $x$ and $y$ matrices in particle/hole space. 
The centered dots indicate matrix multiplication in particle/hole space. We compute the response function 
\begin{equation}
\begin{split}
\frac{\delta \Delta_{m}}{\delta A_\mu(\vec{q}, \tau)}&=\frac{1}{N_c}\llangle \mathcal{T}_\tau \hat{\Delta}_{m} (\vec{q}, \tau) \cJ^p_\mu(\vec{-q},0)  \rrangle \\
&=\frac{U}{N_c} \sum_{\vec{k}\vec{k}'}\sum_{pll'p'} \llangle \mathcal{T}_\tau \psi^\dagger_{p}(\vec{k},\tau) \cdot \mathcal{M}^{m}_{pl} \cdot \psi_{l}(\vec{k}+\vec{q},\tau) \psi_{l'}^{\dagger}(\vec{k}',0) \cdot \cI_{l'p';\mu}(\vec{k}', -\vec{q}) \cdot\psi_{p'}(\vec{k}'-\vec{q},0)  \rrangle
\end{split}
\end{equation}
We have assumed that $\llangle \Delta_{m} (\vec{q})\rrangle = 0$ for $\vec{q}\neq0$ and $\llangle \cJ^p_\mu(\vec{q}=0)  \rrangle =0$ (i.e. the total current vanishes in the groundstate).
The above may be expressed in Matsubara frequency space as
\begin{equation}
\begin{split}
\frac{\delta \Delta_{m}}{\delta A_\mu(\vec{q}, i\omega_n)} 
&= -\frac{U}{\beta N_c} \sum_{pp'll'} \sum_{\vec{k}, ik_{n'}} \Tr \bigg[ \cG_{p' l'}(\vec{k}, ik_{n'}) \cdot \mathcal{M}^{m}_{l'l} \cdot \cG_{lp}(\vec{k}+\vec{q}, ik_{n'} + i\omega_n)\cdot  \cI_{p p';\mu}(\vec{k}+\vec{q}, -\vec{q})\bigg] \\ 
&= -\frac{U}{\beta N_c} \sum_{pp'} \sum_{\vec{k}, ik_{n'}} \Tr \bigg[ \cG_{p' m}(\vec{k}, ik_{n'})  \cdot\tau_- \cdot \cG_{mp}(\vec{k}+\vec{q}, ik_{n'} + i\omega_n) \cdot  \cI_{p p';\mu}(\vec{k}+\vec{q}, -\vec{q})\bigg]
\end{split}
\end{equation}
We follow the steps leading to \eqref{para response zero q}, and we find
\begin{align} \label{delta response}
\frac{\delta\Delta_{m}}{\delta Q_\mu} &= -\frac{U}{N_c}\sum_{\vec{k}}\sum_{ab} \frac{n_F\left(E_a(\vec{k})\right)- n_F\left(E_b(\vec{k})\right)}{E_a(\vec{k}) - E_b(\vec{k})} \bra{\phi_a(\vec{k})} \cM^{m} \ket{\phi_b(\vec{k})}
\bra{\phi_b(\vec{k})}  \cI_\mu(\vec{k},  0) \ket{\phi_a(\vec{k})}\\
&= -\frac{U}{N_c}\sum_{\vec{k}}\sum_{ab} \frac{n_F\left(E_a(\vec{k})\right)- n_F\left(E_b(\vec{k})\right)}{E_a(\vec{k}) - E_b(\vec{k})} \bra{\phi^{m}_a(\vec{k})} \tau_- \ket{\phi^{m}_b(\vec{k})}
\bra{\phi_b(\vec{k})}  \cI_\mu(\vec{k},  0) \ket{\phi_a(\vec{k})}
\end{align}
This is an equation in $\frac{\delta\Delta_{m}}{\delta Q_\mu}$ since this quantity also appears in the right hand side as
\begin{equation} \label{52}
     \cI_\mu(\vec{k},  0) \equiv   \cI^K_\mu(\vec{k}, 0) + \sum_{m} \left(\frac{\delta\Delta_{m}}{\delta Q_\mu}\cM^{m T} + \frac{\delta\Delta^*_{m}}{\delta Q_\mu}\cM^{m} \right)
\end{equation}
where 
\begin{equation}
    \cI^K_\mu(\vec{k}, 0) \equiv \cI^K_\mu(\vec{k}) = \begin{pmatrix} \frac{\partial H}{\partial k_\mu}\big|_{\vec{k}} && 0  \\ 0&& -\frac{\partial H^*}{\partial k_\mu}\big|_{-\vec{k}}\end{pmatrix}
\end{equation}
Plugging this in to the right hand side gives
\begin{align} \label{delta resp}
    \frac{\delta\Delta_{m}}{\delta Q_\mu} = &-\frac{U}{N_c}\sum_{\vec{k}}\sum_{ab} \frac{n_F\left(E_a(\vec{k})\right)- n_F\left(E_b(\vec{k})\right)}{E_a(\vec{k}) - E_b(\vec{k})} \bra{\phi^{m}_a(\vec{k})} \tau_- \ket{\phi^{m}_b(\vec{k})}
\bra{\phi_b(\vec{k})}  \cI^K_\mu(\vec{k}) \ket{\phi_a(\vec{k})} \nonumber\\ 
&- \frac{U}{N_c}\sum_{m'}\frac{\delta\Delta_{m'}}{\delta Q_\mu} \sum_{\vec{k}}\sum_{ab} \frac{n_F\left(E_a(\vec{k})\right)- n_F\left(E_b(\vec{k})\right)}{E_a(\vec{k}) - E_b(\vec{k})} \bra{\phi^{m}_a(\vec{k})} \tau_- \ket{\phi^{m}_b(\vec{k})}
\bra{\phi^{m'}_b(\vec{k})} \tau_+ \ket{\phi^{m'}_a(\vec{k})} \nonumber\\ 
&- \frac{U}{N_c}\sum_{m'}\frac{\delta\Delta^*_{m'}}{\delta Q_\mu} \sum_{\vec{k}}\sum_{ab} \frac{n_F\left(E_a(\vec{k})\right)- n_F\left(E_b(\vec{k})\right)}{E_a(\vec{k}) - E_b(\vec{k})} \bra{\phi^{m}_a(\vec{k})} \tau_- \ket{\phi^{m}_b(\vec{k})}
\bra{\phi^{m'}_b(\vec{k})} \tau_- \ket{\phi^{m'}_a(\vec{k})}
\end{align}
where $\tau_+= \frac{1}{2} (\tau_x + i\tau_y)$. This may be written in the following form
\begin{equation} \label{our equation}
    (C_\mu)_{m} = \sum_{m'} \left(\mathcal{A}_{mm'} \frac{\delta\Delta_{m'}}{\delta Q_\mu} + \mathcal{B}_{mm'} \frac{\delta\Delta^*_{m'}}{\delta Q_\mu} \right)
\end{equation}
where
\begin{equation}\label{ABb}
\begin{split}
\mathcal{A}_{mm'} &=  -\frac{1}{N_c}\sum_{\vec{k}}\sum_{ab} \frac{n_F\left(E_a(\vec{k})\right)- n_F\left(E_b(\vec{k})\right)}{E_a(\vec{k}) - E_b(\vec{k})} \bra{\phi^{m}_a(\vec{k})} \tau_- \ket{\phi^{m}_b(\vec{k})}
\bra{\phi^{m'}_b(\vec{k})} \tau_+ \ket{\phi^{m'}_a(\vec{k})} - \frac{1}{U}\delta_{mm'} \\ 
\mathcal{B}_{mm'} &= -\frac{1}{N_c} \sum_{\vec{k}}\sum_{ab} \frac{n_F\left(E_a(\vec{k})\right)- n_F\left(E_b(\vec{k})\right)}{E_a(\vec{k}) - E_b(\vec{k})} \bra{\phi^{m}_a(\vec{k})} \tau_- \ket{\phi^{m}_b(\vec{k})}
\bra{\phi^{m'}_b(\vec{k})} \tau_- \ket{\phi^{m'}_a(\vec{k})} \\
(C_\mu)_{m} &= \frac{1}{N_c}\sum_{\vec{k}}\sum_{ab} \frac{n_F\left(E_a(\vec{k})\right)- n_F\left(E_b(\vec{k})\right)}{E_a(\vec{k}) - E_b (\vec{k})} \bra{\phi^{m}_a(\vec{k})} \tau_- \ket{\phi^{m}_b(\vec{k})}
\bra{\phi_b(\vec{k})}  \cI^K_\mu(\vec{k}) \ket{\phi_a(\vec{k})}
\end{split}
\end{equation}
 The vector $C_\mu$ corresponds to the first term on the right hand side of \eqref{delta resp}. 

The equation \eqref{our equation} is singular. This follows from gauge invariance. Consider a small rotation of the phase of $\Delta_j$
\begin{equation} \label{goldstone}
\Delta_j \rightarrow e^{i\alpha}\Delta_j \approx (1 + i\alpha)\Delta_j
\end{equation}
where $\alpha\ll 1$. We may think of this as perturbing the Hamiltonian \eqref{mean field hamiltonian} with
\begin{equation} \label{perturbation}
    \hat{\cH}' =  - i\alpha\sum_{j} \Big(\Delta_j c^\dagger_{j,\uparrow}c^\dagger_{j,\downarrow}-\Delta_j^* c_{j,\downarrow}c_{j,\uparrow} \Big)
\end{equation}
We then compute the response of the pairing potential $\Delta_i$ to this static perturbation. On the one hand, the answer is clear, it is
\begin{equation}\label{59}
    \delta\Delta_{m} = i\alpha\Delta_{m}
\end{equation}
On the other hand, the response $\delta\Delta_{m}$ follows from the static linear response formula, which may be written as
\begin{equation}\label{60}
\begin{split}
    \delta\Delta_{m} = &\frac{U}{N_c}\sum_{\vec{k}}\sum_{ab} \frac{n_F\left(E_a(\vec{k})\right)- n_F\left(E_b(\vec{k})\right)}{E_a(\vec{k}) - E_b(\vec{k})} \bra{\phi_a(\vec{k})} \cM^{m} \ket{\phi_b(\vec{k})}
\bra{\phi_b(\vec{k})}\left( -i\alpha\cM^{m T} \Delta_{m}+i\alpha\cM^{m}\Delta^*_{m}\right)\ket{\phi_a(\vec{k})} 
    \\=&U\sum_{m'}\left(\mathcal{A}_{mm'}+\frac{1}{U}\delta_{mm'}\right)(i\alpha\Delta_{m'})-U\sum_{mm'}\mathcal{B}_{mm'}(i\alpha\Delta^*_{m'}) 
    \end{split}
\end{equation}
Combining \eqref{59} and \eqref{60}, we have
\begin{equation} \label{projector}
  \sum_{m'}\left(\mathcal{A}_{mm'}\Delta_{m'}-\mathcal{B}_{mm'}\Delta^*_{m'}\right) = 0
\end{equation}
(This may be viewed as a Ward identity obeyed by the pairing susceptibility.) Thus, for any solution $\frac{\delta\Delta_{\tilde{i}}}{\delta Q_\mu}$ of \eqref{our equation}, we may construct another solution
\begin{equation} \label{eq65}
    \frac{\delta\Delta_{\tilde{i}}}{\delta Q_\mu}\rightarrow\frac{\delta\Delta_{\tilde{i}}}{\delta Q_\mu}'=\frac{\delta\Delta_{\tilde{i}}}{\delta Q_\mu}+i\alpha\Delta_{\tilde{i}} 
\end{equation}
since according to \eqref{projector}, the additional term is ``projected out." The response \eqref{59} corresponds to the Goldstone mode, as is clear from \eqref{goldstone}. The matrix pseudoinverse is convenient for finding a representative solution of a singular equation such as \eqref{our equation}.

%
\subsection{Correction to the Superfluid Weight}
Once $\frac{\delta\Delta_{m}}{\delta Q_\mu}$ is determined, we may compute the correction to the superfluid weight. It is expedient to write
\begin{equation}
\begin{split}
    (D_s)_{\mu\nu} &=  \llangle \hat{T}_{\mu\nu}^{K}(\vec{q}=0,\omega=0)\rrangle +  \frac{1}{N_c}\sum_{\vec{k}}\sum_{ab} \frac{n_F\left(E_a(\vec{k})\right)- n_F\left(E_b(\vec{k})\right)}{E_a(\vec{k}) - E_b(\vec{k})} \bra{\phi_a(\vec{k})}  \cI^K_\mu(\vec{k}, 0) \ket{\phi_b(\vec{k})}
\bra{\phi_b(\vec{k})} \cI_\nu(\vec{k}, 0) \ket{\phi_a(\vec{k})} \\
&=\left(D^{(0)}_s\right)_{\mu\nu} + \frac{1}{N_c}\sum_{\vec{k}}\sum_{ab} \sum_{m}\frac{n_F\left(E_a(\vec{k})\right)- n_F\left(E_b(\vec{k})\right)}{E_a(\vec{k}) - E_b(\vec{k})} \bra{\phi_a(\vec{k})}  \cI^K_\mu(\vec{k}, 0) \ket{\phi_b(\vec{k})}
\bra{\phi_b(\vec{k})} \left(\frac{\delta\Delta_{m}}{\delta Q_\nu}\cM^{m T} + \frac{\delta\Delta^*_{m}}{\delta Q_\nu}\cM^{m} \right) \ket{\phi_a(\vec{k})} \\
&= \left(D^{(0)}_s\right)_{\mu\nu} + 2~\text{Re}\left[\sum_{m}(C_\mu)_{m}\frac{\delta\Delta_{m}^*}{\delta Q_\nu}\right] 
\end{split}
\end{equation}
where 
\begin{equation}
    \left(D^{(0)}_s\right)_{\mu\nu} =  \llangle \hat{T}_{\mu\nu}^{K}(\vec{q}=0,\omega=0)\rrangle  + \Pi^{(0)}_{\mu\nu}(\vec{q}\rightarrow 0, i\omega_n = 0)
\end{equation}
is the usual formula for the superfluid weight and
\begin{equation}
    \Pi^{(0)}_{\mu\nu}(\vec{q}\rightarrow 0, i\omega_n = 0) = \frac{1}{N_c}\sum_{\vec{k}}\sum_{ab} \frac{n_F\left(E_a(\vec{k})\right)- n_F\left(E_b(\vec{k})\right)}{E_a(\vec{k}) - E_b(\vec{k})} \bra{\phi_a(\vec{k})}  \cI^K_\mu(\vec{k}, 0) \ket{\phi_b(\vec{k})}
\bra{\phi_b(\vec{k})} \cI^K_\nu(\vec{k}, 0) \ket{\phi_a(\vec{k})}
\end{equation}
 Therefore the correction to the superfluid weight due to the response of the pairing potential is
\begin{align} \label{delta Dss}
    \left(\delta D_s \right)_{\mu\nu} &= 2~\text{Re}\left[\sum_{m}(C_\mu)_{m}\frac{\delta\Delta_{m}^*}{\delta Q_\nu}\right] \\
    &=\label{the correction}\frac{1}{N_c}\sum_{\vec{k}}\sum_{ab} \sum_{m}\frac{n_F\left(E_a(\vec{k})\right)- n_F\left(E_b(\vec{k})\right)}{E_a(\vec{k}) - E_b(\vec{k})} \bra{\phi_a(\vec{k})}  \cI^K_\mu(\vec{k}, 0) \ket{\phi_b(\vec{k})}
\\&\hspace{30mm} \times  \left( \frac{\delta\Delta_{m}}{\delta Q_\nu}\bra{\phi^{m}_b(\vec{k})}\tau_+ \ket{\phi^{m}_a(\vec{k})} + \frac{\delta\Delta^*_{m}}{\delta Q_\nu}\bra{\phi^{m}_b(\vec{k})}\tau_-\ket{\phi^{m}_a(\vec{k})}  \right)
\end{align}
where we have used the expression for $C_\mu$ is given in \eqref{ABb}. Eq. \eqref{the correction} is the response of the current due to the change in the pairing potential induced by a ``twist" $\vec{Q}$.

\subsubsection{Example: Uniform Superconductor}
We show that $\left(\delta D_s \right)_{\mu\nu}$ vanishes for a uniform $s$-wave system. It is sufficient to show that $C_\mu$ vanishes. The Hamiltonian is
\begin{equation}
    \cH = \sum_{\vec{k}\sigma}\xi_{\vec{k}}c^\dagger_{\vec{k}\sigma}c_{\vec{k}\sigma} - \Delta \sum_{\vec{k}}c^\dagger_{\vec{k}\up}c^\dagger_{-\vec{k}\down}- \Delta^*\sum_{\vec{k}}c_{-\vec{k}\down}c_{\vec{k}\up} \end{equation}
where $V$ is the volume of the system. Let us take $\Delta$ to be real. The Green's function is then
\begin{equation}
    \cG(\vec{k},ik_n) = \frac{ik_n + \xi_{\vec{k}}\tau_3+\Delta\tau_1}{(ik_n)^2 - E_{\vec{k}}^2}
\end{equation}
where $E_{\vec{k}}^2 = \xi_{\vec{k}}^2+\Delta^2$. It is convenient to express $C_\mu$ as
\begin{equation}
\begin{split}
    C_\mu &= -\frac{1}{\beta V}\lim_{\vec{q}\rightarrow0}\sum_{\vec{k}, ik_n}(\partial_\mu\xi_{\vec{k}})\Tr\left[ \tau_- \cdot \cG(\vec{k}+\vec{q}, ik_n)\cdot \cG(\vec{k}, ik_n) \right] \\ &= -\frac{1}{V}\sum_{\vec{k}}(\partial_\mu\xi_{\vec{k}})\frac{\Delta}{2E_{\vec{k}}} \left(n'_F(E_{\vec{k}}) - n'_F(-E_{\vec{k}})\right) \\ &=0
    \end{split}
\end{equation}
where $\partial_\mu = \partial/\partial k_\mu$. The last equality follows from the fact that $n'_F(-E_{\vec{k}}) = n'_F(E_{\vec{k}})$. Thus the correction \eqref{the correction} vanishes, and a spatially non-uniform system is required for the correction to be non-zero.

\subsubsection{Relation to Vertex Corrections}
We briefly sketch how our result may be obtained using current vertex corrections in the Nambu basis. In a superconductor, the photon vertex function \(\Gamma^\mu\) describes the coupling of the system to the electromagnetic vector potential \(A_\mu\). This can be related to the bare vertex \(\gamma^\mu\) (e.g., determined through minimal or Peierls substitution), and the correction arising from interactions encoded in the self-energy \(\Sigma\):

\begin{equation}
\Gamma^\mu - \gamma^\mu = \frac{\delta \Sigma}{\delta A_\mu}.
\end{equation}
In terms of Green’s functions, the electromagnetic kernel is expressed as:
\begin{equation}
\Pi^{\mu\nu}(q) = -i\Tr \left[\int d^4 k \, \gamma^\mu(k+q, k) \cG(k) \, \Gamma^\nu(k, k+q) \, \cG(k+q) \right],
\end{equation}
where \(\cG(k)\) is the full Green’s function in the Nambu basis.
The self-energy \(\Sigma\) may depend on \(A_\mu\), either directly, or indirectly through the order parameter. Indeed, in BCS/BdG theory, the self-energy in the Nambu basis is given by
\begin{equation}
\Sigma(k) = \begin{pmatrix}
0 && \Delta(k) \\
\Delta^\dagger(k) && 0
\end{pmatrix}
\end{equation}
Thus the correction to the electromagnetic kernel is patently given by a term proportional to $\delta \Delta/\delta A_\mu$. One may solve for this correction using the Bethe-Salpeter equation, and it can be shown that this formulation respects gauge invariance via the generalized Ward identity \cite{schrieffer1964}.

\subsubsection{Gauge Invariance}

Physical observables, in particular the correction to the superfluid weight arising from the fluctuations in $\Delta$, must be invariant under global $U(1)$ gauge transformations. This implies that the overall phase of the $\{\Delta_m(\vec{Q})\}$ can vary in an arbitrary way as a function of the probe field $\vec{Q}$. To clarify the point at issue, let us suppose that there exists a choice of the overall phase such that $\{\Delta_m(\vec{Q})\}$ are smooth functions of $\vec{Q}$ in the neighborhood of $\vec{Q}=0$. We may thus approximate
\begin{equation} \label{eq6approx}
\frac{\delta \Delta_m}{\delta Q_{i}} \approx \frac{\Delta_m(\delta Q \unit{e}_i)- \Delta_m(0)}{\delta Q}
\end{equation}
for $\delta Q$ sufficiently small. This approximation for determining $\frac{\delta \Delta_m}{\delta Q_{i}}$ may be most convenient when one has numerical solutions of $\{\Delta_m(\vec{Q})\}$ for various $\vec{Q}$. However, numerical solutions of $\{\Delta_m(\vec{Q})\}$ are not guaranteed to possess a smoothly varying phase as a function of $\vec{Q}$. This is especially salient in systems with vortices, where the singular nature of the phase around vortex cores means that, in general, the phase does not vary smoothly. Let us suppose that the numerical solution $\{\Delta'_m(\vec{Q})\}$ differs  from the smooth solution $\{\Delta_m(\vec{Q})\}$ by a phase: $\Delta'_m(\delta Q \unit{e}_i) = \Delta_m(\delta Q \unit{e}_i)e^{i\theta(\delta Q \unit{e}_i)}$ and $\Delta'_m(0) = \Delta_m(0)$. Then, using the approximation Eq. \eqref{eq6approx}, we would find 
\begin{align}
\frac{\delta \Delta'_m}{\delta Q_{i}} \approx \frac{\Delta'_m(\delta Q \unit{e}_i)- \Delta'_m(0)}{\delta Q} &= \frac{\Delta_m(\delta Q \unit{e}_i)e^{i\theta(\delta Q \unit{e}_i)}- \Delta_m(0)}{\delta Q} \\&\approx \left(\frac{e^{i\theta(\delta Q \unit{e}_i)} - 1}{\delta Q} \right)\Delta_m(0) + \frac{\delta \Delta_m}{\delta Q_{i}} \label{eq8}
\end{align}
Again, since smoothness is not guaranteed, $e^{i\theta(\delta Q \unit{e}_i)} - 1$ need not be small. Therefore, in order that the result for $\delta D_s$ be gauge invariant, it should return the same result whether $\frac{\delta \Delta'_m}{\delta Q_{i}}$ or $\frac{\delta \Delta_m}{\delta Q_{i}}$ is substituted into the formula. Equivalently, it should ``project out" terms of the form $z_i\Delta_m(0)$ where $z_i=\frac{e^{i\theta(\delta Q \unit{e}_i)} - 1}{\delta Q} $ is a (large) complex number. 

We note that this is more restrictive than standard discussions of gauge invariance, since we allow arbitrarily rapid phase variations in the order parameter. This is essential for computing $\delta \Delta_m / \delta Q_i$ via finite difference approximation. In contrast, solutions to Eq. \eqref{our equation} are defined modulo a constant corresponding to an imaginary $z_i$ (see Eq. \eqref{eq65}). This corresponds to the collective mode which is projected out, according to the generalized Ward identity (see previous section), thus ensuring gauge invariance. However, when using the finite difference approximation, both real and imaginary components of $z_i$ must be projected out, as we have argued in the previous paragraph. We will argue that Eq. \eqref{delta Dss} is gauge-invariant in this latter sense. As we have mentioned below Eq. \eqref{delta Dss}, $ \delta D_s$ is determined by the response of the current due to the change in the pairing potential. In other words, it is determined by the crossed susceptibility $\chi_{J, \Delta}$. Using this identification of the correction to the superfluid stiffness with the current-pairing susceptibility, we may see why it is gauge invariant according to the discussion above, provided scaling the pairing potential by a complex number produces no total current response in equilibrium. Consider an inhomogeneous superconductor described by a position-dependent pairing potential \(\Delta(\vec{r})\). The current density \(\vec{J}(\vec{r})\) in such a system is given by
\begin{equation}
\vec{J}(\vec{r}) \propto \text{Im} \left( \Delta^*(\vec{r}) (\nabla + 2i \vec{A}) \Delta(\vec{r}) \right),
\end{equation}
where \(\Delta(\vec{r}) = |\Delta(\vec{r})| e^{i \theta(\vec{r})}\) is the complex pairing potential. In equilibrium, the total current is zero:
\begin{equation}
\vec{J}_{\text{tot}}=\int \vec{J}(\vec{r}) \, d\vec{r} = 0,
\end{equation} We now scale the pairing potential by a complex number \( z = |z| e^{i \phi} \). The scaled pairing potential is
\begin{equation}
\Delta'(\vec{r}) = z \Delta(\vec{r}) = |z| e^{i \phi} \Delta(\vec{r}) = |z| |\Delta(\vec{r})| e^{i (\theta(\vec{r}) + \phi)}.
\end{equation}
To determine the effect of this scaling on the current density, we compute the new current density \(\vec{J}'(\vec{r})\) for the scaled pairing potential:
\begin{equation}
\vec{J}'(\vec{r}) \propto \text{Im} \left( (\Delta'(\vec{r}))^* (\nabla + 2i \vec{A}) \Delta'(\vec{r}) \right).
\end{equation} Therefore, the new current density is
\begin{equation}
\vec{J}'(\vec{r}) \propto \text{Im} \left( |z|^2 \Delta^*(\vec{r}) (\nabla + 2i \vec{A}) \Delta(\vec{r}) \right) = |z|^2 \text{Im} \left( \Delta^*(\vec{r}) (\nabla + 2i \vec{A}) \Delta(\vec{r}) \right) = |z|^2 \vec{J}(\vec{r}).
\end{equation} Since \(\vec{J}_{\text{tot}}=0\), it follows that
\begin{equation}
\vec{J}'_{\text{tot}} = |z|^2 \vec{J}_{\text{tot}} = 0.
\end{equation} Thus, the change in the total current (the induced current response) due to scaling the pairing potential is
\begin{equation}
\delta \vec{J}_{\text{tot}} = 0.
\end{equation}

We have verified numerically that Eq. \eqref{delta Dss} projects out $\frac{\delta \Delta_m}{\delta Q_{i}} =  z_i \Delta_m$, and that it gives the same result under arbitrary variations in the phase $\theta(\vec{q})$, in the ground state for all cases studied in this work, thus achieving gauge invariance.

%

\begin{figure}
    \centering
    \includegraphics[width=\linewidth]{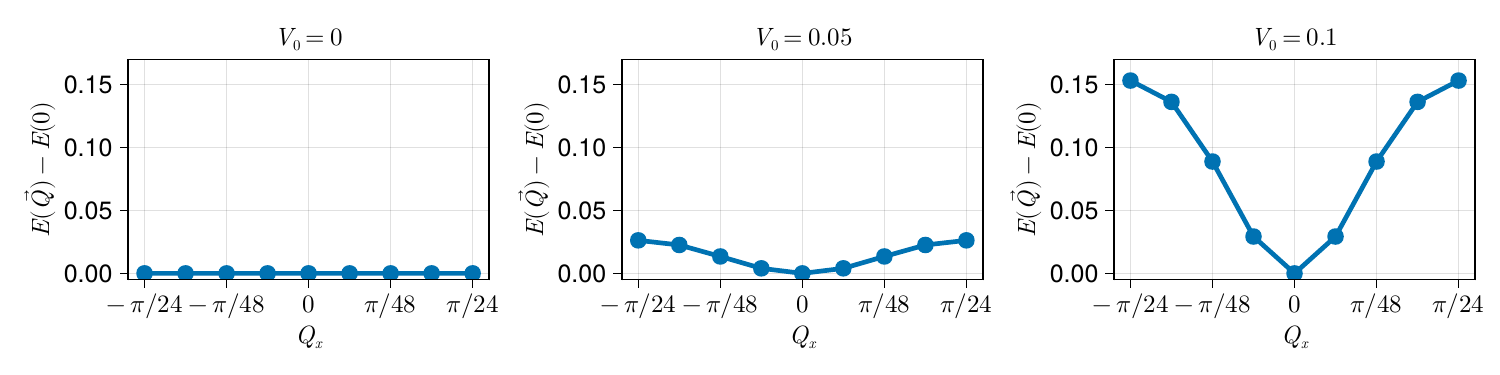}
    \caption{The energy as a function of $Q_x$ for the system corresponding to Fig. 2 of the main text. The cyan points in that figure are computed by taking a numerical second derivative at the minimum of the energy using $Q$ points very close to that value (not shown here). Without a periodic potential ($V_0 = 0$), the system forms a vortex lattice that freely moves as $Q$ is varied. Thus, the energy does not depend on $Q$, as seen in the left panel, and the superfluid weight vanishes. When a periodic potential is applied, the vortices become somewhat obstructed (pinned), resulting in the reemergence of superfluid weight (center and right panels).}
    \label{fig:enter-label}
\end{figure}

\begin{figure}
    \centering
    \includegraphics[width=.75\linewidth]{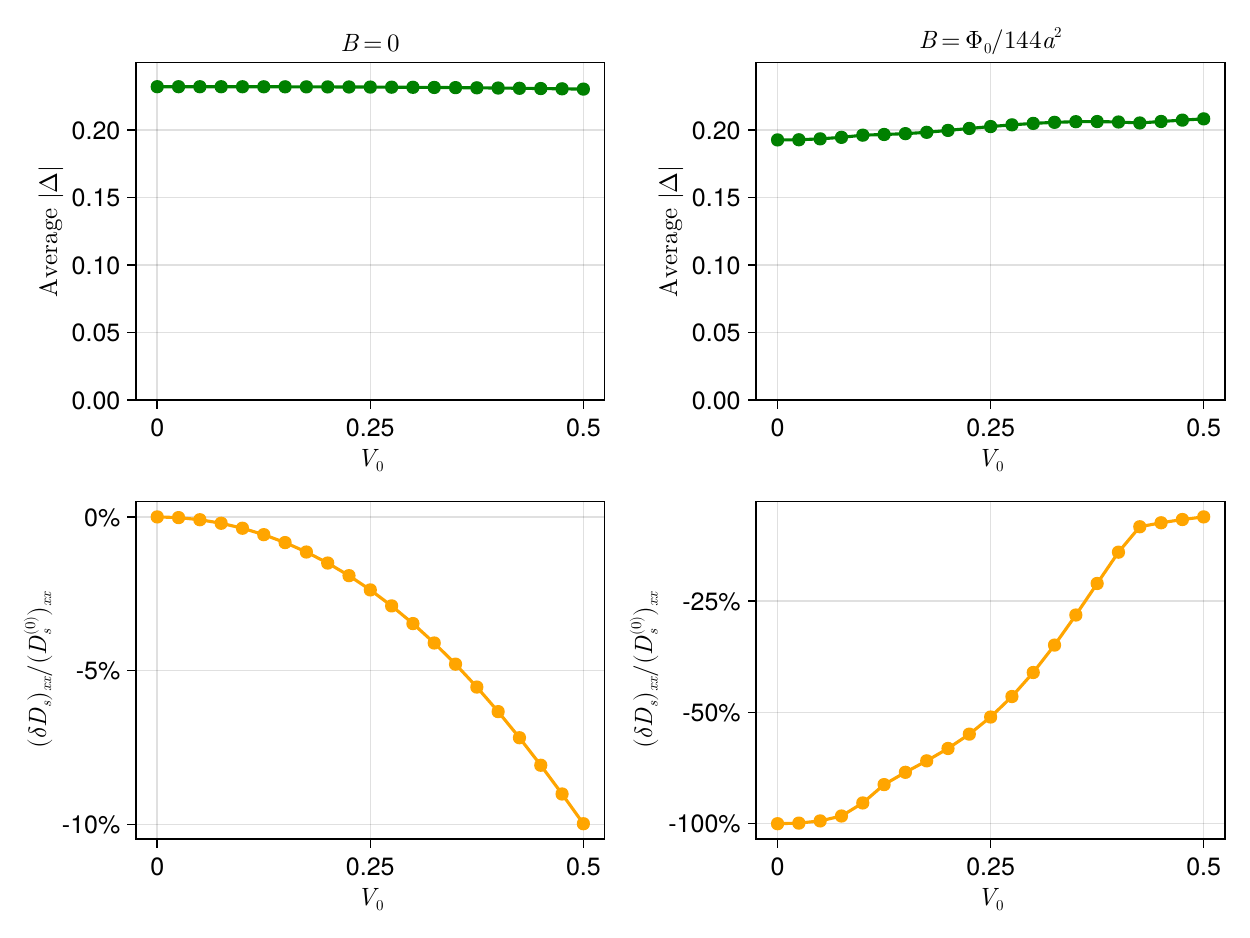}
    \caption{Top row: the spatially averaged value of $|\Delta|$ as a function of the strength of the applied potential without (left) and with (right) an applied magnetic field. Areas with high potential $V(\vec{r})$ tend to reduce $|\Delta|$, while areas with low potential tend to enhance it, resulting in $|\Delta|$ roughly maintaining its average as $V_0$ increases. Bottom Row: The correction to the superfluid weight due to the response of $\Delta$, expressed as a percentage of $D^{(0)}_s$, the uncorrected superfluid weight, without (left) and with (right) an applied magnetic field. Under an applied magnetic field, a vortex lattice forms, and if the vortices are unpinned ($V_0=0$), the correction completely eliminates the superfluid weight, which is incorrectly given as non-zero according to $D^{(0)}_s$. This is seen from the $-100 \%$ correction shown in the bottom right panel at $V_0 = 0$.}
    \label{fig:enter-label}
\end{figure}

\clearpage

\section{Inversion of Equation~\ceq{eq.dD2}}
As stated in the main text the square matrix $K$ that appears in Eq.~\ceq{eq.dD2} is singular, it has rank one less than its dimension, $n$, 
reflecting the fact that the superconducting order parameter is defined
apart from an overall phase factor.
To obtain $\delta \Delta_m/{\delta A_\mu}\big|_0$ we need to calculate the pseudoinverse, $\tilde K$, of  $K$.
To do this we first perform a single-value-decomposition (SVD) of $K$:
\be
 K = U \Lambda V^\dagger
\ee
where $U$ and $V$ are $n\times n$ unitary matrices and $\Lambda$ is a $n\times n$ diagonal matrix of the form
\be
 \Lambda = 
 \begin{pmatrix}  
   \lambda_1   &   0        & \cdots     & \cdots     & 0 \\
   0           & \lambda_2  & 0          & \cdots     & 0 \\
   \vdots      &   0        & \lambda_3  & 0          & 0 \\
   \vdots      & \vdots     & \vdots     & \ddots     & \vdots \\
   0           &   	0       &    0       &   0        & \lambda_n=0  
  \end{pmatrix}.
\ee
Let
\be
 \tilde\Lambda = 
 \begin{pmatrix}  
   \lambda_1^{-1}   &   0        & \cdots     & \cdots     & 0 \\
   0           & \lambda_2^{-1}  & 0          & \cdots     & 0 \\
   \vdots      &   0        & \lambda_3^{-1}  & 0          & 0 \\
   \vdots      & \vdots     & \vdots     & \ddots     & \vdots \\
   0           &   	0       &    0       &   0        & \tilde\lambda_n=0  
  \end{pmatrix}.
\ee
then, the pseudoinverse of $K$ is given by
\be
 \tilde K = V \tilde\Lambda U^\dagger
\ee
and 
\be
 \left.\frac{\delta\Delta_m}{\delta A_\mu}\right|_0= \tilde K C_\mu + (1-\tilde K K)W
\ee
where $C_\mu$ is the column vector with elements $\left\{(C_\mu)_m\right\}$, and
$W$ is the vector containing the free parameters. In our case, only the last element
of the diagonal matrix $(1-\tilde K K)$ is nonzero leaving just one free parameter,
corresponding to the overall gauge phase factor.

\section{Superconductor with vortex lattice}
We consider a superconductor with a 2D vortex lattice in the $(x,y)$ plane
induced by a background perpendicular magnetic field $B_z$. 
The presence of this field is taken into account within the tight model Hamiltonian~\ceq{mean field hamiltonian} 
via the introduction of a Peierls phase.
This has the effect of altering the translation group of the underlying square lattice to that 
of the magnetic translation group \cite{Hofstadter1976}. 
A magnetic unit cell must be chosen such that an integer number of magnetic flux quanta $\Phi_0 = h/e$ 
due to $B_z$ thread the 2D systems. We chose the magnetic unit cells to be $M a\times M a$ with one flux quantum threading it.

\end{document}